\documentclass[12pt,preprint]{aastex}

\usepackage{amsmath}
\usepackage{amssymb}

\newcommand{\be}{\begin{equation}}
\newcommand{\ee}{\end{equation}}
\newcommand{\bea}{\begin{eqnarray}}
\newcommand{\eea}{\end{eqnarray}}

\newcommand{\ec}{{\cal E}}

\shorttitle{Milky Way Models}
\shortauthors{Widrow & Pym}

\begin{document}

\title{Dynamical Blueprints for Galaxies}

\author{Lawrence M. Widrow\altaffilmark{1}}
\affil{Department of Physics, Engineering Physics, and 
Astronomy, Queen's University, 
Kingston, ON, K7L 3N6, Canada}
\altaffiltext{1}{widrow@astro.queensu.ca}

\author{Brent Pym\altaffilmark{2}}
\affil{Department of Mathematics, University of Toronto,
40 St. George Street, Toronto, ON, M5S 2E4, Canada}
\altaffiltext{2}{bpym@math.toronto.edu}

\author{John Dubinski\altaffilmark{3}}
\affil{Department of Astronomy and Astrophysics, University of 
Toronto, 60 St. George Street, Toronto, ON, M5S 3H8, Canada
\\ \medskip
(submitted to the Astrophysical Journal, October, 2007)}
\altaffiltext{3}{dubinski@astro.utoronto.ca}

\bigskip

\begin{abstract}

  We present an axisymmetric, equilibrium model for late-type galaxies
  which consists of an exponential disk, a Sersic bulge, and a cuspy
  dark halo.  The model is specified by a phase space distribution
  function which, in turn, depends on the integrals of motion.
  Bayesian statistics and the Markov Chain Monte Carlo method are used
  to tailor the model to satisfy observational data and theoretical
  constraints.  By way of example, we construct a chain of $10^5$
  models for the Milky Way designed to fit a wide range of photometric
  and kinematic observations.  From this chain, we calculate the
  probability distribution function of important Galactic parameters
  such as the Sersic index of the bulge, the disk scale length, and
  the disk, bulge, and halo masses.  We also calculate the probability
  distribution function of the local dark matter velocity dispersion
  and density, two quantities of paramount significance for
  terrestrial dark matter detection experiments.
  
  Though the Milky Way models in our chain all satisfy the prescribed
  observational constraints, they vary considerably in key structural
  parameters and therefore respond differently to non-axisymmetric
  perturbations.  We simulate the evolution of twenty-five models
  which have different Toomre $Q$ and Goldreich-Tremaine $X$
  parameters.  Virtually all of these models form a bar, though some,
  more quickly than others.  The bar pattern speeds are $\sim 40 -
  50\,{\rm km\,s^{-1}\,kpc^{-1}}$ at the time when they form and then
  decrease, presumably due to coupling of the bar with the halo.
  Since the Galactic bar has a pattern speed $\sim 50\,{\rm km\,s^{-1}
    \,kpc^{-1}}$ we conclude that it must have formed recently.

\end{abstract}



\section{INTRODUCTION}

Dynamical galactic models serve a variety of purposes.  They may be
used to interpret the structural and kinematical observables of
galaxies -- surface brightness profiles, rotation curves and velocity
dispersion profiles -- in terms of intrinsic three-dimensional density
and velocity distributions.  Dynamical models also provide a starting
point for controlled simulations of complicated processes such as the
formation of bars and spiral structure.  In short, galactic modeling
provides the essential interface between observations and detailed
theories of galaxy formation.


In this paper, we introduce a new dynamical model for late-type
galaxies which comprises a disk, bulge, and dark halo.  The model is
derived from equilibrium solutions to the collisionless Boltzmann and
Poisson equations.  It extremely flexible and may be tailored to
satisfy observational data and theoretical constraints.  We use
Bayesian statistics and the Markov Chain Monte Carlo (MCMC) method to
implement these constraints and to determine the probability
distribution function (PDF) of the model in the full multi-dimensional
parameter space.

Our model builds upon earlier work by \citet{kui95} and \citet{wid05}.
The original Kuijken \& Dubinski model consists of an exponential
disk, a King-model bulge, and a lowered Evans-model halo and has the
attractive feature that the phase space distribution function (DF) is
built from analytic functions of the integrals of motion.  No
additional assumptions about the velocity-space distribution are made.
By contrast, the widely-used approach described in \citet{her93} (see
also \citet{springel99}) assumes that the local velocity distributions
of the halo and bulge particles are Gaussian with dispersions
estimated from the Jeans equations.  This approach leads to models
which are slighly out of equilibrium.  When used as initial conditions
in N-body experiments, they readjust to a different state from the one
proposed (see, for example, \citet{kaz04}).

There are two main disadvantages of the Kuijken and Dubinski models.
First, the bulge and halo have constant density (or weakly cuspy)
centers whereas actual bulges and dark halos may have central density
cusps.  Second, the structure of the bulge and halo are determined
{\it implicitly} by the model parameters.  (By construction, the
disk's structural parameters, namely its radial and vertical scale
lengths, its mass, and its truncation radius, are determined
explicitly by the model parameters.)  \citet{wid05} built a galactic
model with $r^{-1}$-density cusps for both the bulge and halo but
again, with a DF that determines the structure of the bulge and halo
implicitly.

For our new model, the closed-form DF is abandoned in favor of a
numerical DF which is designed to yield, to a very good approximation,
user-specified density profiles for the bulge and halo.  That is, the
density profiles of the bulge and halo are now {\it explicit}
functions of the model parameters.  The present version of the model
allows for a Sersic bulge and a halo with $\rho\propto r^{-\gamma}$ as
$r\to 0$ where $\gamma$ is between $0$ and $2$.

Our model is specified in terms of fifteen or so parameters.  How are
these parameters selected?  One approach is to choose models at random
and identify the ones that satisfy certain general constraints (e.g.,
the Tully-Fisher and size-luminosity relations).  The result would be
a catalog of disk-bulge-halo systems which could be used to study
kinematical and dynamical trends such as the circular speed-central
velocity dispersion ($V_c-\sigma_0$) relation (see, for example,
\citet{cou07}).  A catalog of this type could also be used for N-body
studies of mergers or interactions between galaxies in a cosmological
environment.  A second approach, and the one pursued here, is to build
models for specific galaxies.  Following \citet{kui95} and
\citet{wid05}, we use the Milky Way as our illustrative example.

Modeling the Milky Way is a time-honoured endeavor; notable examples
include \citet{inn73}, \citet{clu77}, \citet{bs80}, \citet{cal81},
\citet{kui91}, \citet{roh88}, \citet{mal95}; \citet{koc96},
\citet{eva99}, and \citet{kly02}.  Our construction of dynamical Milky
Way models is in the spirit of the mass model survey by
\citet{deh98}.  Their models comprise a multi-component disk, a bulge,
and a halo and are characterized by ten parameters.  Twenty-two
examples are presented, each the result of a maximum likelihood
analysis in which some parameters are held fixed while others are
allowed to vary.  The Dehnen \& Binney likelihood function is
constructed by comparing model predictions with six sets of
observational data: the inner and outer Galactic rotation curves, the
Oort constants, the mass at large radii, the local vertical force, and
the line-of-sight velocity dispersion in Baade's window.  
The advantage of our models is that they not only describe the
potential-density pair for the Galaxy but also, the underlying
DF.  We therefore have the ability to examine the stability of our
Galactic models using N-body simulations, an issue that is often
ignored (but see \citet{sel85} and \citet{fux97}).  

For the most part, we adopt Dehnen \& Binney's choice of observational
data though we include more complete observations of the line-of-sight
dispersion in the bulge region as well as photometric data from the
COBE satellite.  We also present what we believe to be a more balanced
treatment of the likelihood function.  Most significantly, we bring to
the problem the powerful tools of Bayesian statistics and MCMC.  These
tools allow us to map out PDFs of both input parameters and derived
quantities.

Though our model represents an axisymmetric, equilibrium system, it is
susceptible to non-axisymmetric instabilities and therefore provides a
natural starting point for numerical studies of galactic dynamics.  An
N-body realization of the model can be easily generated from the DF
and then used as the initial conditions for a numerical simulation.

The Milky Way models in our MCMC series all satisfy the observational
constraints but vary considerably in their structural properties.  We
simulate a selection of twenty-five models which span a wide range in
Toomre $Q$ \citep{toomre64} and Goldreich-Tremaine $X$ \citep{gt78,
  gt79} parameters and find that a bar develops in virtually all of
the cases.  The onset of the bar instability can occur immediately or
after several Gyr, depending on the model.

We present the model in Section 2, review the observational
constraints in Section 3, and provide a summary of the essentials of
Bayesian statistics and the MCMC method in Section 4.  We discuss some
preliminaries including our choice of prior probabilities, in Section
5.  We present the results of our MCMC analysis in Section 6 and the
results of our bar formation simulations in Section 7.  In Section 8
we summarize our main conclusions and speculate on how we might
improve upon and extend the models and MCMC analysis.

\section{GALACTIC MODELS}

We consider axisymmetric, collisionless systems whose DF is of the
form
\begin{equation}\label{eq:totalDF}
f\left (\ec,\,L_z,\,E_z\right )~=~
f_{\rm disk}\left (\ec,\,L_z,\,E_z\right )~+~
f_{\rm bulge}\left (\ec\right )
~+~f_{\rm halo}\left (\ec\right )
\end{equation}
where $\ec\equiv -E$ is the relative energy, $L_z$ is the angular
momentum about the symmetry axis, and $E_z$ is the energy associated
with vertical motions of stars in the disk \citep{kui95,wid05}.  For
time-independent, axisymmetric systems ${\cal E}$ and $L_z$ are
integrals of motion while $E_z$ is an approximate integral of motion
for disk stars on nearly circular orbits.  Jeans theorem implies that
a system generated by equation \ref{eq:totalDF} will be in approximate
equilibrium.

Integrating equation\,\ref{eq:totalDF} over all velocities yields the
density in terms of the gravitational potential, $\Phi$, and the
cylindrical coordinates $R$ and $z$:
\begin{equation}\label{eq:totalrho}
\rho\left (R,\,z,\,\Psi\right )~=~
\rho_{\rm disk}\left (R,\,z,\,\Psi\right )~+~
\rho_{\rm bulge}\left (\Psi\right )
~+~\rho_{\rm halo}\left (\Psi\right )
\end{equation}
where $\Psi\equiv -\Phi$ is the relative potential.  (Note that
implicit in equation \ref{eq:totalDF} is the assumption that the bulge
and halo velocity dispersions are isotropic.)  Self-consistency
requires that $\rho$ and $\Psi$ satisfy Poisson's equation
\begin{equation}\label{poisson}
\nabla^2\Psi ~=~-
4\pi\rho\left (R,\,z,\,\Psi\right )
\end{equation}
which is accomplished, in practice, through an iterative scheme.
(Note that here and throughout, we set Newton's constant $G=1$.)

\citet{kui95} chose $f_{\rm bulge}$ to be the King model DF
\citep{kin66} and $f_{\rm halo}$ to be the lowered Evans model DF from
\citet{kui94}.  (The latter depends on $L_z$ as well as $\ec$ thereby
allowing for flattened halos.)  Their models have two main
shortcomings.  First, bulges and halos may have central density cusps
whereas the King and lowered-Evans DFs yield density profiles with
constant-density cores.  Second, the relationship between the model
parameters and the density profiles of the bulge and halo is implicit
rather than explicit and not particularly intuitive.

\citet{wid05} built galactic models with cuspy ($\rho\propto r^{-1}$
as $r\to 0$) bulges and halos.  Specifically, they chose the
\citet{her90} DF for the bulge and a DF from \citet{wid00} for the
halo.  The latter was constructed to yield the so-called NFW profile
\begin{equation}\label{eq:nfwprofile}
\rho_{\rm NFW}(r) = \frac{\rho_h}{\left (r/a_h\right )\left (1 +
r/a_h\right )^2}
\end{equation}
\citep{nfw96}.  \citet{wid05} altered these DFs in an attempt to account
for the gravitational potential of the other components though the
modifications were somewhat {\it ad hoc}.  Consequently, the density
profile of the halo differed from the NFW profile and likewise for the
bulge.

A further drawback of the \citet{wid05} model is that the Hernquist
bulge and NFW halo are arguably too restrictive.  The Hernquist DF
yields a system whose surface density profile is well approximated by
the $R^{1/4}$-law \citep{deV48}.  The bulges of late-type galaxies are
found to have surface brightness profiles which follow the more
general Sersic law,
\begin{equation}\label{eq:sersic}
\Sigma(r) = \Sigma_0 e^{-b\left (R/R_e\right )^{1/n}}~,
\end{equation}
with Sersic index $n$ between $0.6$ and $2$ \citep{and95, cou96}.
Likewise, dark matter halos may have density profiles more general
than the NFW form.  Since the work of \citet{nfw96}, there has been
considerable debate over the actual form of halo density profiles.
\citet{moo99} find evidence in their simulations for steeper cusps
($\rho\propto r^{-1.5}$).  More recently, \citet{nav04} conclude that
the logarithmic slope of the halo density profiles decreases steadily
with radius though their results are still consistent with equation
\ref{eq:nfwprofile}.  On the observational front, the rotation curves
of dark matter-dominated low surface brightness galaxies appear to
favour constant density cores \citep{moo94, flo94, mcg98, vdB00}
though this interpretation of the data is being challenged on a number
of fronts.

For our new models, we begin by choosing target density profiles,
$\tilde{\rho}_{\rm bulge}$ and $\tilde{\rho}_{\rm halo}$, for the
bulge and halo.  Assume, for the moment, that the system is
spherically symmetric.  Through the Abel integral transform,
\begin{equation}\label{eq:abel}
f_i\left (\ec\right ) = \frac{1}{\sqrt{8}\pi^2}
\int_0^\ec \frac{d^2\tilde{\rho}_i}{d\Psi_{\rm total}^2}
\frac{d\Psi_{\rm total}}{\sqrt{\ec - \Psi_{\rm total}}} ~~~~~~~~~~
i ~=~ {\rm bulge~or~halo}~,
\end{equation}
\citep{BT}, we can construct bulge and halo DFs which yield the target
density profiles.  In the case of an isolated halo or bulge,
$\Psi_{\rm total}$ is the potential derived from $\tilde\rho_{\rm
  halo}$ or $\tilde\rho_{\rm bulge}$ and equation\,\ref{eq:abel}
reduces to the usual expression for the DF of a spherically symmetric
system with isotropic velocities.  The DF for a system following the
Sersic law was found with this method by \citet{cio91}.  DFs for NFW
halos were found by \citet{zha97}, \citet{wid00}, and \citet{lok00}.
For a composite system or one with an external potential, one simply
replaces the $\Psi$ derived from $\tilde\rho_i$ with the total
gravitational potential.  \citet{tre02} used this method to derive DFs
for bulges with central black holes by setting $\Psi_{\rm total} =
\Psi_{\rm bulge} + GM_{\rm black hole}/r$.

Equation\,\ref{eq:abel} is only valid for spherically symmetric
systems, a condition violated once a disk is included.  Our approach
is to use a spherical approximation (essentially, the monopole term of
a spherical harmonics expansion) for the disk potential in evaluating
$\Psi_{\rm total}$.  We stress that equation\,\ref{eq:abel} is used to
construct $f_{\rm halo}(\ec)$ and $f_{\rm bulge}(\ec)$, not to solve
for $\Psi(R,z)$ and $\rho(R,z)$.  We can use $f_{\rm halo}(\ec)$ and
$f_{\rm bulge}(\ec)$ in equation \ref{eq:totalDF} even though the
composite system is not spherically symmetric; A DF of the form $f =
f(\ec)$ yields an equilibrium system in {\it any} time-independent
potential regardless of the spatial symmetries of the
potential.\footnote{A {\it self-consistent} system with a DF that
  depends only on the energy must be spherically symmetric \citep{BT}.
  The statement in the text applies to systems in which there is an
  external potential that does not necessarily respect spherical
  symmetry.  Here the potential due to the disk plays the role of an
  external potential to the halo and bulge.}  The bulge and halo of
the final model are axisymmetric (but not spherically symmetric) since
isodensity surfaces follow isopotential surfaces (equation
\ref{eq:totalrho}), the latter being flattened by the disk.  As we
demonstrate below, the spherically-averaged density profiles of the
bulge and halo are very close to the target profiles.

\subsection{\it Target Density Profiles}

We choose the target density profile for the bulge to be
\begin{equation}\label{eq:prugnielsimien}
\tilde{\rho}_{\rm bulge}(r) = \rho_b\left (\frac{r}{R_e}\right )^{-p} 
e^{-b\left (r/R_e\right )^{1/n}}
\end{equation}
which yields the Sersic law (equation \ref{eq:sersic}) 
for the projected surface density profile provided one sets $p = 1 -
0.6097/n + 0.05563/n^2$ \citep{ps97,ter05}.  Note that here and in
equation \ref{eq:sersic}, $\Sigma_0$, $R_e$ and $n$ are free
parameters while the constant $b$ is adjusted so that $R_e$ encloses
half the total projected light or mass.  In our models, we use
\begin{equation}\label{vbdef}
\sigma_b\equiv
\left (4\pi n b^{n(p-2)} \Gamma\left (n\left (2-p\right )\right )
R_e^2 \rho_b \right )^{1/2}
\end{equation}
rather than $\rho_b$ to parametrize the overall density scale of the
bulge models.  With this definition, $\sigma_b^2$ corresponds to the
depth of the gravitational potential associated with the bulge.

We choose the target density profile of the halo to be
\begin{equation}\label{eq:haloprofile}
\tilde{\rho}_{\rm halo} = \frac{2^{2-\gamma}\sigma_h^2}{4\pi a_h^2}
\frac{1}{\left (r/a_h\right )^{\gamma}
\left (1 + r/a_h\right )^{3-\gamma}}\,C\left (r;r_h,\delta r_h\right )
\end{equation}
where $C$ is a truncation function that smoothly goes from unity to
zero at $r=r_h$ over a width $\delta r_h$.  We use the function
\begin{equation}
C\left (r;r_h,\delta r_h\right ) = \frac{1}{2}
{\rm erfc}\left (\frac{r-r_h}{\sqrt{2}\delta r_h}\right )~.
\end{equation}
The halo profile is therefore characterized by five parameters: $r_h$,
$\delta r_h$, the halo scale length, $a_h$, the velocity scale,
$\sigma_h$, and the central ``cusp strength'', $\gamma$.  For $r<r_h$,
the mass interior to radius $r$ is given by

\begin{equation}
M(r) = 2^{2-\gamma}\sigma_h^2 a_h\left (
\frac{1}{1+r/a_h} + \log\left (1+r/a_h\right )\right )~.
\end{equation}

Following \citet{kui95}, we adjust the disk's DF so that its space
density falls off approximately exponentially in $R$ and as $\rm
sech^2$ in $z$ with radial and vertical scale lengths $R_d$ and $z_d$
respectively.  The disk is truncated at a radius $R_{\rm out}$ with a
truncation sharpness parameter $\delta R_d$.  In addition, we choose a
DF where the radial dispersion profile is approximately exponential:
\begin{equation}\label{eq:radialdispersion}
\sigma_R^2(R) = \sigma_{R0}^2 \exp{\left (-R/R_\sigma\right )}~.
\end{equation}
For simplicity, we set $R_\sigma = R_d$ in accord with observations by
\citet{bot93}.

\section{OBSERVATIONAL CONSTRAINTS}

We use nine sets of observational data as constraints on our
Milky Way models.  Five of these data sets -- the inner and outer rotation
curves, the Oort constants, the vertical force in the solar
neighborhood, and the total mass at large radii -- are largely taken
from \citet{deh98} and references therein.  We incorporate
measurements of the line-of-sight bulge dispersion from the
compilation of data by \citet{tre02} as well as estimates of the local
velocity ellipsoid from \citet{bin98}.  We also use dust-corrected
near-infrared DIRBE data from the COBE satellite \citep{bin97}.

\begin{itemize}

\item{\it Inner rotation curve}

Inside the solar circle, the Galactic rotation curve is usually
presented in terms of the ``terminal velocity'', the peak velocity
along a given line-of-sight at Galactic coordinates $b=0$ and
$|l|<\pi/2$.  Assuming that the Galaxy is axisymmetric, $v_{\rm term}$
is given by

\begin{equation}\label{eq:vterm}
v_{\rm term} = v_c(R)-v_c\left (R_0\right )\sin{l}
\end{equation}

\noindent where $R_0$ is the distance from the Sun to the Galactic
center and $v_c$ is the circular speed (see, for example,
\citet{bin98}).  Following \citet{deh98} we use data from HI emission
observations by \citet{mal95} restricted to the range $\sin{l}\ge
0.3$ so as to avoid distortions from the bar.

\item {\it Outer rotation curve}

The radial velocity of an object 
relative to the local standard of rest, $v_{\rm lsr}$, is
related to the circular rotation curve through the equation
\begin{equation}\label{eq:outerRC}
v_{\rm lsr}=\left (
\frac{R_0}{R}v_c\left (R\right )-
v_c\left (R_0\right )\right )\cos{b}\sin{l}
\end{equation}
where $R=\left (d^2\cos^2{b}+R_0^2-2R_0d\cos{b}\sin{l}\right )^{1/2}$,
$(l,\,b)$ are the Galactic coordinates, and $d$ is the distance to the
object.  Measurements of $v_{{\rm lsr},i}$ and $d_i$ are compared to
model estimates for $W(R)$ and $d(R)$ where $W(R)\equiv \left
  (R_0/R\right )v_c(R)-v_c(R_0)\equiv v_{lsr}/\cos{b}\sin{l}$.  $R$ is
regarded as a free parameter which is adjusted to minimize
\begin{equation}\label{eq:chi_i}
\chi^2_i = \left (\frac{W(R)-W_i}{\Delta W_i}\right )^2+
\left (\frac{d(R)-d_i}{\Delta d_i}\right )^2
\end{equation}
where $W_i\equiv v_{{\rm lsr},i}/\cos{b}\sin{l}$.  In what follows we
use data from \citet{bra93} with the same restrictions as in
\citet{deh98} (i.e., $l\le 155^\circ$ or $l\ge 205^\circ$, $d>1\,{\rm
  kpc}$, and $W<0$) so as to avoid contamination from non-circular
motions.

\item {\it Local circular speed}

  A further constraint from the rotation curve of the Galaxy comes
  from estimates of the circular speed at the solar radius,
  $v_c(R_0)$.  Here, we adopt the estimate of \citet{rei99} who
  carried out VLBA observations of Sgr $A^*$.  Under the assumption
  that Sgr $A^*$ is at the center of the Galaxy, they found
\begin{equation}\label{eq:vcirc}
v_c(R_0) = \left (219\pm 20\, {\rm km\,s^{-1}}\right )\left (
\frac{R_0}{8\,{\rm kpc}}\right )
\end{equation}
which is consistent with other other estimates \citep{sac97}.

\item {\it Vertical force above the disk}

\citet{kui91} use K dwarf stars as tracers of the gravitational
potential above the Galactic plane thereby placing a constraint on the
total surface density in the solar neighborhood.  They find

\begin{equation}\label{eq:kuigil1}
\frac{|K_z\left (1.1\,{\rm kpc}\right )|}{2\pi G}
= 71\pm 6\,M_\odot {\rm pc}^{-2}
\end{equation}

\noindent independent of the relative contributions of the disk and
halo.  By including additional constraints on the local circular
speed, Galactocentric distance of the Sun, and Oort constants, one can
ferret out the separate contributions of the disk and halo to the
local surface density.  Doing so, \citet{kui91} found

\begin{equation}\label{eq:kuigil2}
\Sigma_{\rm disk}
= 48\pm 9\,M_\odot {\rm pc}^{-2}~,
\end{equation}

\noindent in excellent agreement with estimates of known
matter in the solar neighborhood.  We adopt equation \ref{eq:kuigil1}
as the constraint on the vertical force at $\left (R,\,z\right ) =
\left (R_0,\,1.1\,{\rm kpc}\right )$ and equation \ref{eq:kuigil2} the
constraint on the surface density of the disk at $R=R_0$.

\item {\it Oort constants}

The Oort constants,
\begin{equation}\label{eq:oort_a}
A \equiv \frac{1}{2}
	\left (\frac{v_c}{R}-\frac{\partial v_c}{\partial R}\right )_{R=R_0}
~~~~~~~{\rm and}~~~~~~~~
B \equiv -\frac{1}{2}
	\left (\frac{v_c}{R}+\frac{\partial v_c}{\partial R}\right )_{R=R_0}~,
\end{equation}
measure, respectively, the local shear and vorticity in the Galactic
disk.  Here we adopt the constraints
\begin{equation}\label{eq:oort_contraints}
A = 14.8\pm 0.8\,{\rm km\,s^{-1}\,kpc^{-1}}~~~~~~~
B = 12.4\pm 0.6\,{\rm km\,s^{-1}\,kpc^{-1}}
\end{equation}
from the \citet{feast97} analysis of Cepheid proper motion
measurements by the Hipparcos satellite.

\item {\it Local velocity ellipsoid}

  The kinematics of stars in the solar neighborhood provide important
  constraints on the structure of the Milky Way.  The observation that
  $\overline{v_R^2}\ne \overline{v_z^2}$ already tells us that the
  disk DF necessarily involves a third integral of motion
  \citep{bin87}.  Our constraints for the local velocity ellipsoid are
  taken from Table 10.4 of \citet{bin98} which in turn were derived
  from \citet{edv93}.  \citet{bin98} give separate values for the thin
  and thick disks.  Since our models assume a single disk component we
  use a mass-weighted average \citep{wid05}.

\item {\it Bulge dispersion}

  Observations of the line-of-sight velocity dispersion in the
  direction of the bulge provide important constraints on the bulge
  parameter and, to a lesser extent, the parameters of the other
  components.  We use measurements of the line-of-sight dispersion
  between $4\,{\rm pc}$ and $1300\,{\rm pc}$ from the compilation by
  \citet{tre02}.  Since the bulge is triaxial, the measured
  line-of-sight dispersion depends on the observer's orientation to its
  principal axes.  Our line-of-sight to the Galactic center is 
  approximately $20^\circ$ from the long axis of the bulge
  \citep{bin97} and therefore the measured line-of-sight dispersion
  will be systematically higher than the values one would obtain
  assuming a spherical bulge.  Following \citet{tre02}, we adjust the
  measured dispersions downward by a factor 1.07 to account for this
  effect.

\item {\it Mass at large radii}

  The observed velocity distribution of the Milky Way satellite system
  and the dynamics of the Magellanic Stream, together with
  measurements of the high-velocity tail of the local stellar velocity
  distribution, provide constraints on the large-scale mass
  distribution of the Galactic halo.  Following \citet{deh98}, who
  base their arguments on analyses by \citet{koc96} and \citet{lin95}, we
  adopt

\begin{equation}\label{eq:mass100} 
M\left (r<100\,{\rm kpc}\right ) = \left (7\pm 2.5\right
)\times 10^{11} \,M_\odot 
\end{equation}

\noindent as a constraint on the mass distribution at large radii.

\item{Surface photometry}

  The distribution of stars in the Milky Way is most easily determined
  from observations in the near infrared where stellar emission
  dominates over dust emission.  Though dust is more transparent at
  these wavelengths than in the optical, extinction due to dust is
  still significant toward the Galactic center.  \citet{spe96}
  produced extinction-corrected maps of the inner Galaxy based on the
  DIRBE data set and a three-dimensional dust model (see also
  \citet{freud}).  \citet{bin97} constructed three-dimensional models
  for the light distribution of the disk and bulge based on these maps
  with the aim of constraining the structural parameters of the
  Galactic bar.

  Our initial goal is to construct axisymmetric models for the Galaxy.
  Toward this end, we use the surface brightness as a function of
  $l$ at mid Galactic-latitudes ($3^\circ < |b| <
  4.5^\circ$) from \citet{bin97} (their Figure 2, lower panel) where
  the effects of the bar are not so pronounced (i.e., where their
  axisymmetric model adequately reproduces the observed surface
  brightness profile).

\end{itemize}

The mass model survey of \citet{deh98} employs a maximum likelihood
analysis where the likelihood function is $\exp\left (-\chi^2_{\rm
    DB}\right )$ with
\begin{equation}\label{eq:pseudochi2}
\chi^2_{\rm DB} = 
\frac{W_{\rm in}}{N_{\rm in}}\chi^2_{\rm in}+
\frac{W_{\rm out}}{N_{\rm out}}\chi^2_{\rm out}+
\frac{W_{\rm other}}{N_{\rm other}}\chi^2_{\rm other}~.
\end{equation}
The subscripts ``in'', ``out'', and ``other'' refer to the inner and
outer rotation curve constraints and the other constraints (e.g., Oort
constants, vertical force) respectively.  The $N_i$ are the numbers of
data points actually used while $W_i$ are weights introduced by
\citet{deh98} to account for ``the number of really independent
constraints''.  That is, the $W_i$ are meant to compensate for the
fact that a quantity such as the Oort constant $A$ has been obtained
from a large number of data points.  \citet{deh98} choose $W_{\rm in}
= W_{\rm out} = W_{\rm other} = 6$ though admit that the choice of
$W_i$'s are ``subject to ones prejudices''.

In our view, the likelihood function should be $\exp{\left (-\chi_{\rm
tot}^2/2\right )}$ where
\begin{equation}\label{chi2total}
\chi_{\rm tot}^2 = \chi_{\rm in}^2 + \chi_{\rm out}^2 + \chi_{\rm
other}^2~.
\end{equation}
The fact that the Oort constant constraints are obtained from a large
number of (raw) data points is already accounted for by the small
quoted errors.  Dividing $\chi^2_{\rm in}$ by $N_{\rm in}$ unfairly
short-changes the rotation curve data in favour of the Oort constant
constraints, and so forth.

To survey the model parameter space, \citet{deh98} adopt the following
approach; fix certain parameters and maximize the likelihood function
by allowing the remaining parameters to vary.  The procedure is then
repeated with the fixed parameters set to different values, or
different subsets of parameters held fixed.  The result is a rather
uneven survey of the full parameter space.  A similar exercise was
carried out by \citet{wid03} for M31 using the original Kuijken and
Dubinski models together with rotation curve, velocity dispersion, and
surface brightness data.  This procedure was also used for both M31
and the Milky Way in \citet{wid05}.  Bayesian statistics and MCMC
provide a more complete picture of the model parameter space as we now
demonstrate.

\section{BAYESIAN ANALYSIS AND MCMC}

Our aim is to calculate the posterior probability density function,
$p(M|D,I)$, of a Galactic model $M$ given data $D$ and prior
information $I$.  $M$ is specified by the fifteen model parameters as
well as additional astronomical parameters -- here $R_0$ and the
mass-to-light ratios of the disk and bulge, $\left (M/L\right )_d$ and
$\left (M/L\right )_b$.  We collect the parameters into a vector ${\bf
  A}$ with components $A^j$ where $j=1..N$ and $N$ is the total number
of parameters.  From Bayes' theorem
\begin{equation}\label{eq:bayes}
p(M|D,I) = \frac{p(M|I)p(D|M,I)}{p(D|I)}
\end{equation}
where $p(M|I)$ is the prior probability density, $p(D|M,I)$ is the
likelihood function, and $p(D|I)\equiv \int \, p(M|D,I)\,d{\bf A}$ is
a normalization factor.  Our choice of priors is described in the next
section.

MCMC is an efficient means of calculating $p(M|D,I)$ whereby one
constructs a sequence or ``chain'' of models whose density in
parameter space is proportional to the posterior PDF provided the
chain is long enough to have fully explored all ``important'' regions
of parameter space.  Marginalization -- that is, integration over a
subset of parameters -- is trivial; simply project the chain onto the
appropriate subspace and compute the density of points.  Likewise, the
PDF of any model-dependent quantity is obtained by making a histogram
of the quantity over the chain of models.

Our Markov chain is constructed via the Metropolis-Hastings algorithm
\citep{met53, hast70} as outlined in \citet{gre05}.  The chain of
models is described by the sequence ${\bf A}_i,~i=0,1,2...$~.  We
begin with a starting point ${\bf A}_0$.  A candidate for ${\bf A}_1$
is chosen according to the {\it jumping rule} (also known as the {\it
proposal distribution}), $q\left ({\bf A}_1| {\bf A}_0\right )$.  ${\bf
A}_1$ is accepted with probability equal to ${\rm min}\left
\{1,\,r\right \}$ where
\begin{equation}
\label{eq:ratio}
r\equiv
\frac{p\left ({\bf A}_1|D,I\right )}{p\left ({\bf A}_0|D,I\right )}
\frac{q\left ({\bf A}_0|{\bf A}_1\right )}
{q\left ({\bf A}_1|{\bf A}_0\right )}~.
\end{equation}
Otherwise, ${\bf A}_1$ is set equal to ${\bf A}_0$.  The process is
then repeated for ${\bf A}_2$.

The success of an MCMC analysis rests, by and large, on choosing a
suitable jumping rule.  If the step size from ${\bf A}_n$ to a
candidate for ${\bf A}_{n+1}$ is too small, the chain will move slowly
through parameter space.  On the other hand, if the step size is too
large, most attempts to find a new point in parameter space will fail.
In either case, exploration of parameter space, often referred to as
mixing, can require a prohibitively large amount of computing
resources.  Ideally, the jumping rule is shaped like the PDF but
scaled by a factor $2.4/\sqrt{N}$ \citep{gel95} which explains why it
is often referred to as the proposal distribution.

In this work, we take $q$ to be a multivariate Gaussian 
so that
\begin{equation}\label{eq:proposal}
{\bf A}_{n+1} = {\bf A}_n + {\bf D}\cdot {\bf G}
\end{equation}
where ${\bf G}$ is a vector of length $N$ whose components are
Gaussian random variables with unit variance and ${\bf D}$ is a
user-specified $N\times N$ matrix.  Since neither ${\bf D}$ nor ${\bf
G}$ depend on the model parameters, $q\left ({\bf A}_{n+1}| {\bf
A}_{n}\right ) = q\left ({\bf A}_{n}| {\bf A}_{n+1}\right )$ and
therefore $r = p\left ({\bf A}_{n+1}|D,I\right )/p\left ({\bf
A}_n|D,I\right )$

We begin with a simple ansatz for the proposal distribution in which
${\bf D}$ is a diagonal matrix whose non-zero elements are given by
our best guess for the variance of each of the model parameters
multiplied by $2.4/\sqrt{N}$.  From a short chain of a few thousand
models we estimate the covariance matrix
\begin{equation}\label{covariance}
C_{ij} = \frac{\langle \left (A^i-\bar A^i\right ) \left (A^j-\bar
A^j\right )\rangle} {\langle A^i\rangle\langle A^j\rangle} 
\end{equation}
where $\langle \dots\rangle$ denotes an average along the chain.  Our
improved expression for the proposal distribution is given by
equation\,\ref{eq:proposal} with ${\bf D^2} = \left (2.4^2/N\right
)\,{\bf C}$.

Each ``data point'' carries with it a quoted error.  Of course, the
observer may have underestimated the error or there may be features in
the data which cannot be explained by the model.  Both situations can
be handled by introducing an unknown error parameter for each data set
which is added in quadrature to the quoted error \citep{gre05}.  For
the purpose of the MCMC calculation, these error parameters are simply
incorporated into an expanded definition of ${\bf A}$, that is,
treated as model parameters.

\section{PRELIMINARIES AND PRIORS}

Simple arguments, based on general features of the Galaxy, provide
preliminary estimates for the model parameters which in turn guide our
choices of the prior probabilities used in the MCMC analysis.  We
assume that the priors for each of the model parameters are non-zero
over a range somewhat larger than the range suggested by these
estimates.  For parameters that correspond to a physical scale (e.g.,
$R_e$, $M_d$, $v_h$) we assume a Jeffrey prior, essentially, equal
probability in logarithmic intervals over the prescribed range.  For
dimensionless parameters, such as the halo cusp strength and Sersic
index, we assume a flat prior.  (See \citet{gre05} for a discussion.)

The projected velocity dispersion profile toward the Galactic bulge,
$\sigma_p(R)$, reaches a peak value of $\sim 130\,{\rm km\, s^{-1}}$
at a radius $\sim 200\,{\rm pc}$ \citep{tre02}.  On the other hand,
estimates of the half-light or effective radius of the bulge, $R_e$,
range from $570\,{\rm pc}$ to $920\,{\rm pc}$ (see \citet{tre02} and
references therein).

\begin{figure}
\epsscale{1.0}
\plotone{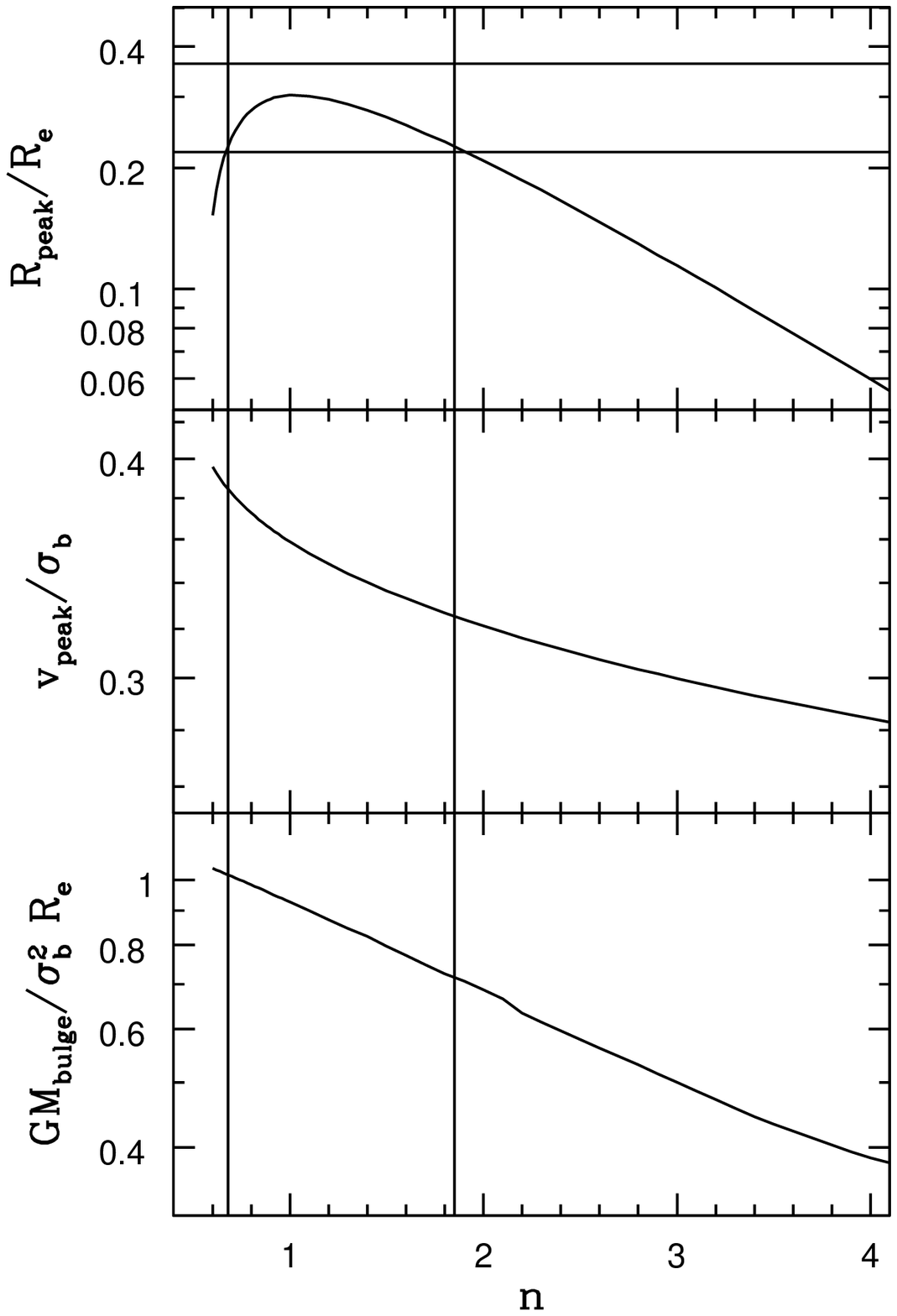}
\caption{Characteristics of the line-of-sight velocity dispersion
  profile for the \citet{ps97} density profile as a function of Sersic
  index $n$.  The curve $\sigma_p(R)$ rises to a peak value of $v_{\rm
    peak}$ at a projected radius $R_{\rm peak}$.  Shown are the
  dimensionless ratios $R_{\rm peak}/R_e$ (upper panel), $v_{\rm
    peak}/\sigma_b$ (middle panel), and $GM_{\rm bulge}/\sigma_b^2
  R_e$ (lower panel) as functions of $n$.  Data for the Galaxy
  compiled by \citet{tre02} suggest that $R_{\rm peak}/R_e$ lies in
  the range $0.22-0.36$ (horizontal lines).  Vertical lines then
  delineate range of values for $n$.}
\label{fig:sersic}
\end{figure}

The projected velocity dispersion of the \citet{ps97} profile exhibits
a similar structure to that of the Milky Way: $\sigma_p(R)$ is
non-zero at $R=0$, rises to a peak value of $\sigma_{\rm peak}$ at a
radius $R_{\rm peak}$ and then decreases with radius (see Figure 10 of
\citet{ps97} as well as earlier work by \citet{bin80} and
\citet{cio91}).  The dimensionless ratios $R_{\rm peak}/R_e$,
$\sigma_{\rm peak}/\sigma_b$ and $M_{\rm bulge}/\sigma_b^2 R_e$ are
functions of $n$ as shown in Figure \ref{fig:sersic}.  From the figure
we deduce that for the Milky Way, $n$ is less than $2$, $R_e$ is
between $0.57\, {\rm kpc}$ and $0.92\, {\rm kpc}$, $\sigma_b$ is
between $340\,{\rm km \, s^{-1}}$ and $400\,{\rm km \,s^{-1}}$, and
$M_b$ is between $1\times 10^{10}\,M_{\odot}-3.4\times
10^{10}\,M_\odot$.

\citet{bin97} constructed a model for the luminosity density of the
Galaxy to fit data from the DIRBE experiment.  Their model consisted of
a triaxial bulge and double exponential disk with bulge-to-total
luminosity ratio of $0.16$.  Subsequently, \citet{mal95b} derived a
total L-band luminosity for the Milky Way of $7.1\times
10^{10}\,L_{\odot}$ with $1.1\times 10^{10}\,L_\odot$ 
attributed to the bulge.

Stellar population synthesis models provide estimates for the stellar
mass-to-light ratios in different wavebands (\citet{bell01} and
references therein).  The L-band stellar mass-to-light ratio for the
disk is expected to be in the range $0.5-0.65$ in solar units
\citep{deJ07} assuming a scaled Salpeter IMF (\citet{bell01}) 
and the Pegase population synthesis model.
The mass-to-light ratio for the bulge could be somewhat higher.  On
the other hand, since our model does not include a separate gas disk,
the effective mass-to-light ratio for the disk must account for any
gas and should therefore be higher than the value for a pure stellar
disk.  The local stellar-to-gas ratio is $\sim 1.6$ (see Table 1 of
\citet{BT}) and therefore the effective $\left (M/L\right )_d$ might
be closer to $1$.  Together with our estimate for the disk luminosity,
we conclude that $M_d$ is between $3\times 10^{10}$ and $6\times
10^{10}\,M_\odot$.

\citet{rei93} reviewed estimates of the distance from the Sun to the
Galactic center and concluded that $R_0 = 8.0\pm 0.5\,{\rm
  kpc}$.  More recently \citet{eis03} observed the star S2 in orbit
about the Galaxy's massive central black hole using the ESO VLT and
found $R_0 = 7.94\pm 0.42\,{\rm kpc}$.

\citet{sac97} reviewed estimates of the radial scale length of the
Galactic disk and found $R_d = 3.0\pm 1\,{\rm kpc}$.  More recent
estimates show a similarly large spread in values.  \citet{zhe01}
found $R_d = 2.75\pm 0.3\,{\rm kpc}$ from HST observations of M dwarfs
while \citet{lop02} found $R_d = 3.3^{+0.5}_{-0.4}\,{\rm kpc}$ from an
analysis of old stellar populations using 2MASS survey data.  As
emphasized by \citet{sac97}, the ratio $R_0/R_d$ is observationally
more secure than $R_d$.  The estimates collected in her review show
$R_0/R_d$ between $2.7$ and $3.5$.

The disk scale height parameter, $h_d$, is more difficult to constrain
since the Galactic disk comprises at least three distinct components,
the gas disk, the thin disk, and the thick disk whereas our model has
a single-component disk.  Multi-component disks will be incorporated
into future versions of the model but for the time being, $h_d$ must
represent the vertical mass distribution of all disk-like components.
From \citet{sac97} we surmise that $h_d$ is between $0.2$ and $1\,{\rm
  kpc}$.

The radial velocity dispersion in the solar neighborhood is $36\pm
5.4\,{\rm km\,s^{-1}}$ \citep{bin98}.  Together with estimates of the
radial scale length of the disk and with equation
\ref{eq:radialdispersion}, this translates into a range of possible
values for $\sigma_{R0}$.

We allow the halo parameters to vary over a wide range of values.  For
example, we assume that the prior probability distribution of $\gamma$
is uniform between $0$ and $1.5$ and non-zero otherwise.  An
alternative approach is to use cosmological models of halo and galaxy
formation to guide ones choice of the halo parameters (see, for
example, \citet{val03}), but given uncertainties in the exact nature
of adiabatic compression, variations among halo profiles found in
different simulations, and possible discrepancies between halo
profiles as inferred from observations and those found in simulations,
we take a more conservative approach.  Furthermore, since the data do
not probe the Galaxy beyond $100\,{\rm kpc}$ we set $r_h=100\,{\rm
  kpc}$ and $\delta r_h = 5\,{\rm kpc}$ though no physical meaning
should be ascribed th these values; $r_h$ can be increased without changing
the fit to the data.

Our choices for the prior probabilities of the model parameters are
found in Table 1.  In addition to the parameters listed in Table 1, we
include unknown error parameters (see previous section) for the
terminal velocity, outer rotation curve, bulge dispersion, and surface
brightness profile data sets.  Thus, our MCMC analysis is run with
seventeen parameters.

\section{MCMC RESULTS}

\begin{figure}
\epsscale{1.0}
\plotone{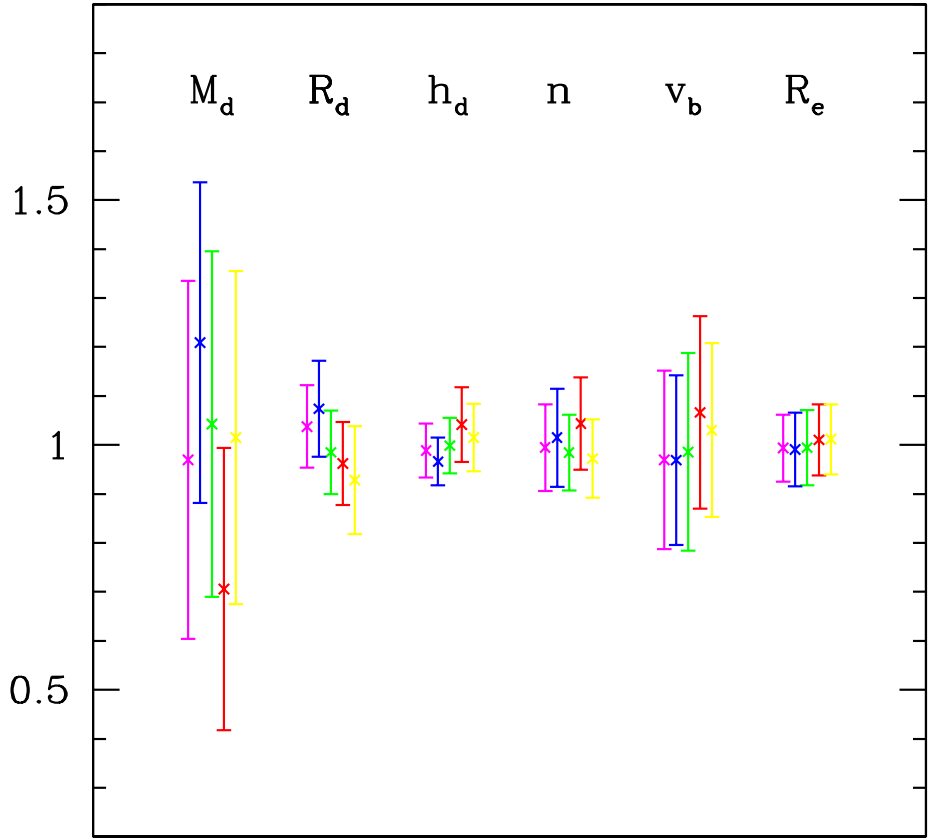}
\caption{Average values and $1\sigma$ error bars for a selection of
  six model parameters as calculated from each of the five separate
  Markov chains.  Values of the parameters are normalized by
  dividing by the overall average.}
\label{fig:mixing}
\end{figure}

We generate five Markov chains of length $2\times 10^4$.  Each chain
has roughly $3400$ distinct members corresponding to an acceptance
rate of $17\%$.  We can calculate the PDFs for different quantities
using the five chains individually and in combination.  If mixing has
been achieved then the results will be the same to within statistical
uncertainties.  In Figure \ref{fig:mixing} we illustrate that this is
indeed the case by plotting the average values with $1\sigma$
error-bars for the a selection of six model parameters calculated for
each of the five chains.  For convenience, the values are normalized
by dividing by the overall average.

\subsection{\it Selected Models}

\begin{figure}
\epsscale{1.0}
\plotone{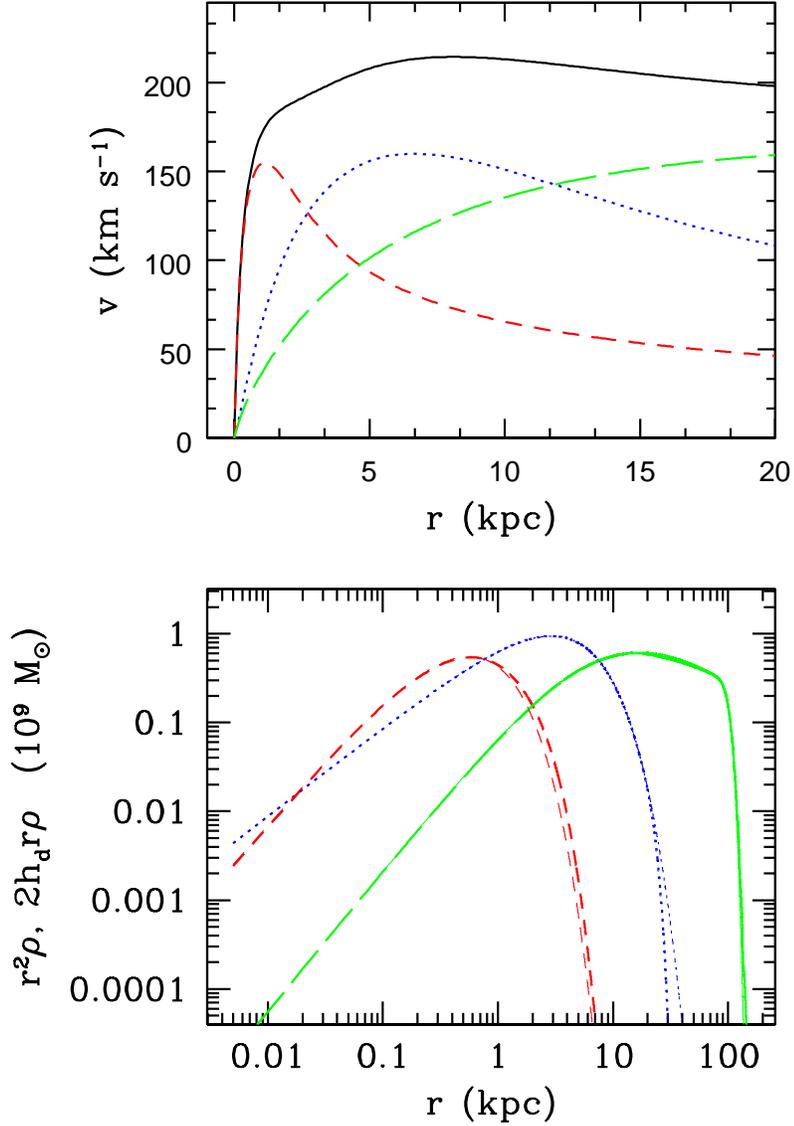}
\caption{Rotation curve (top panel) and density profile (bottom) as a
  function of $r$ for a model with $n\simeq 1$ and $\gamma\simeq 1$.
  Red (dashed) line is for the bulge; blue (dotted) line is for the
  disk; green (long-dashed) line is for the halo.  In the top panel,
  the solid (black) line shows the total rotation curve.  In the lower
  panel, we plot $r^2\rho$ for the bulge and halo and $h_d r\rho$ for
  the disk, quantities proportional to the mass in radial bins.  Also
  shown in the lower panel (thin lines) are the ``target'' density
  profiles for the bulge and halo (equations \ref{eq:prugnielsimien} and
  \ref{eq:nfwprofile} respectively).}
\label{fig:samplegalaxy}
\end{figure}

In Figure \ref{fig:samplegalaxy} we show the rotation curve and
density profile for a model from our MCMC series with $n\simeq 1$ and
$\gamma\simeq 1$.  Also shown are the target bulge and halo profiles
(equations \ref{eq:prugnielsimien} and \ref{eq:haloprofile}).  We see
that the model profiles are very close to the target ones.  By
comparison the halo profiles in the Widrow \& Dubinski models differ
significantly from the target (NFW) profile (see Figure 2, top panel, of
\citet{wid05}).

We display results for the surface brightness profile (Figure 4), the
terminal velocity (Figure 5), and the bulge line-of-sight velocity
dispersion (Figure 6) for three models from our MCMC analysis.  We
choose models with $\gamma\simeq 1$ and Sersic parameters $n \simeq
0.6,\,1$ and $2$.  We also include a model with $n=4$, that is, with
essentially an $R^{1/4}$-law bulge.  Since our MCMC run does not find
any models with a Sersic parameter this large we generate this model
by fixing $n=4$ and allowing the remaining parameters to vary.

While suitable models are found for $n$ between $0.6$ and $2$ this is
not the case for $n=4$.  This result is in agreement with studies of
bulges in late-type spiral galaxies \citep{and95, cou96, mac03} and
suggests that the Galaxy has a pseudo-bulge rather than a
classical bulge (see \citet{kormendy04} for a review).

The models clearly have a difficulty reproducing the shape of the
line-of-sight velocity dispersion profile.  In particular, the local
velocity minimum found in the data at $R\simeq 3\,{\rm pc}$ is much
deeper than is allowed by the models.  The discrepancy may indicate
that the density profile of the Galactic bulge is significantly
different from the one proposed by \citet{ps97} or that velocity-space
anisotropy and deviations from spherical symmetry are necessary to
model the dispersion profile in the innermost region of the bulge
\citep{spe96, fux97, freud}.

\subsection{\it Statistical Overview}

The maximum {\it a posterior} values and boundaries of the 68.3\%
credibility regions for the input parameters and calculated quantities
are given in Tables 2 and 3 respectively.  Not surprisingly, $a_h$ and
$\gamma$ exhibit the largest fractional uncertainties.  Most of the
data used in this study pertains to the region of the Galaxy near and
inside the Sun's orbit about the Galactic center; Equation
\ref{eq:mass100} provides a rather weak constraint on the mass at
large radii (or equivalently, the circular speed at these radii).  We
note that $a_h$ and $\gamma$ are correlated in the sense that models
with larger values of $a_h$ have $\gamma$ closer to $1$ --- large
constant-density cores ($\gamma\simeq 0$ and $a_h\simeq 20-30\,{\rm
  kpc}$ are disfavored by the data.

In Figure 7 we show the PDFs for the disk and bulge masses as well as
the halo mass within $10\,{\rm kpc}$ and $100\,{\rm kpc}$.  We also
show the pseudo-likelihood function for the twenty-two models
considered in \citet{deh98}.  Specifically, we plot
\begin{equation}\label{eq:DB}
{\cal L}_{DB} = e^{-\left (\chi_{\rm DB}^2 - \chi_{\rm min}^2\right )}
\end{equation} 
where $\chi_{DB}^2$ is from their Table 3 with $\chi_{\rm min}^2$ set
to the value for their best-fit model.  Our results for the disk,
bulge, and halo masses are consistent with those of \citet{deh98}.
Figure \ref{fig:masspdfs} illustrates the advantages of the MCMC
method.  The Dehnen \& Binney analysis involves twenty-two separate
maximum-likelihood calculations characterized by the authors' {\it ad
  hoc} choices of fixed and free parameters.  MCMC, on the other hand,
yields the full multi-dimensional posterior probability function
from which PDFs for one or more parameters are easily obtained.

Figure \ref{fig:ro_rd} provides a contour plot of the PDF in
the $R_0-R_d$ plane.  Our results are consistent with previous 
estimates of these two quantities.  The plot also shows the general
trend that models with higher values of $R_d$ tend to favor higher
values of $R_0$.

In Figure 9 we show the PDFs for $n$ and $\gamma$.  As noted above,
our analysis clearly favours bulges with surface brightness profiles
closer to an exponential than to de Vaucouleurs $R^{1/4}$-law.  Our
results allow for a dark halo with a wide range of inner logarithmic
density slopes which includes the NFW value as well as steeper and
shallower indices.

\subsection{{\it Comparison with Published Milky Way Models}}

In Figure \ref{fig:modelcomparison}, we compare values for the disk
and bulge masses from our Markov chain analysis with those from a
number of popular Milky Way models.  One of the earliest mass models
was constructed by \citet{bss83}.  While they focus on constraining
the Galactic spheroid through star counts (see also \citet{bs80}) they
also fit the Galactic rotation curve by modeling the mass
distribution of the disk, bulge, and dark halo.  Their choice for the
disk and bulge masses ($M_d = 5.6\times 10^{10}\, M_\odot$ and $M_b =
1.1\times 10^{10}\,M_\odot$) is represented in Figure
\ref{fig:modelcomparison} by the solid triangle.

From Shuttle IRT observations, \citet{kent91} derived total K-band
luminosities for the disk and bulge of $L_d = 4.9\times
10^{10}\,L_\odot$ and $L_b = 1.1\times 10^{10}\,L_\odot$,
respectively.  (Note that these values differ slightly from the values
quoted in the original paper because a different value for the K
magnitude of the Sun is used.  \citet{ken92} went on to construct
disk-bulge-halo mass models based on these results and was able to fit
the Galactic rotation curve for three choices of the disk
mass-to-light ratio: a maximal disk model ($\left (M/L\right )_d =
1.3$), a high disk model ($\left (M/L\right )_d = 1.0$), and a low
disk model ($\left (M/L\right )_d = 0.68$).  In each case, the bulge
mass-to-light ratio was $\left (M/L\right )_b = 1.0$.  Kent's models
are shown in Figure \ref{fig:modelcomparison} as blue, red, and green
stars.

\citet{kly02} constructed models for the Milky Way and Andromeda
galaxies based on cosmologically motivated theories of disk formation.
Their favored model for the Milky Way has $M_d = 4.0\times 10^{10}\,
M_\odot$ and $M_b = 0.8\times 10^{10}\,M_\odot$ and is represented in
Figure \ref{fig:modelcomparison} by the solid square.

Since its discovery \citep{ibata94, ibata95}, the Sagittarius dwarf
galaxy has held the promise of providing useful constraints on the
size and shape of the Milky Way's dark halo.  The usual approach is to
simulate the tidal disruption of Sagittarius as it passes through the
Galactic potential and compare the distribution of tidal debris with
photometric and kinematic observations of the observed tidal streams
(see \citet{law05} for a recent example and references to earlier
work).  \citet{johnston99} introduced a model for the Galactic
potential which has now become a standard for work in this area.  The
\citet{johnston99} values, $M_d =1.0\times 10^{11}\, M_\odot$ and $M_b
= 3.4\times 10^{10}\,M_\odot$, are shown in Figure
\ref{fig:modelcomparison} as an open square.

Our models occupy a smaller region of $M_d-M_b$ parameter space than
is spanned by popular models from the literature.  The
\citet{johnston99} choices for $M_d$ and $M_b$ are inconsistent with
our results by factors of $2.5$ and $4$, respectively.  More to the
point, their choices fall well outside the region of acceptable
models.  The choices for disk and bulge masses in \citet{bss83} and
\citet{ken92} are consistent with our results as is
the preferred model from \citet{kly02}.

\subsection{{\it Implications for Dark Matter Detection Experiments}}

Models of the Milky Way are an essential ingredient in the analysis of
dark matter detection experiments.  For example, microlensing
experiments, which attempt to measure the distribution of massive compact
halo objects (MACHOs) along various lines-of-sight through the halo,
require a model for the MACHO component of the halo as well as
the distribution of stars in the disk.

In analyzing their 5.7 year data set, the MACHO experiment considered
a wide range of Galactic models \citep{alcock00}.  Here we focus on
their ``standard model''.  This model includes both a thin and thick
disk, each with radial scale length $R_d = 4\,{\rm kpc}$ and total
disk mass is $M_d = 4.5\times 10^{10}\,{\rm M}_\odot$.  The halo is
modeled as a cored isothermal sphere with a density profile given by
\begin{equation}\label{eq:machohalo}
\rho(r) = \rho_0\frac{a^2 + R_0^2}{a^2 + r^2}
\end{equation}
where, for their standard model, \citet{alcock00} assumed
$R_0=8.5\,{\rm kpc}$, $a = 5\,{\rm kpc}$ and $\rho_0 = 0.0079\,{\rm
  M}_\odot {\rm pc}^{-3}$.  Though they did not model the bulge
explicitly, we can infer its mass through the requirement that the
circular rotation speed of the Galaxy at $r=R_0$ is $\simeq 220\,{\rm
  km s}^{-1}$.  Doing so yields $M_b\simeq 2.6\times 10^{10}\,{\rm
  M}_\odot$ for their standard model (open triangle in Figure
\ref{fig:modelcomparison}).  Evidently, the bulge mass is inconsistent
with our results by a factor of 3.

Terrestrial dark matter detection experiments aim to observe
elementary particle dark matter candidates (e.g., WIMPs or axions) as
they interact with a detector on Earth.  These experiments are
therefore sensitive to the local density and velocity distribution of
dark matter particles.  Estimates for the local dark matter density
range from $0.005-0.02\,{\rm M}_\odot {\rm pc}^{-3}$ ( $0.2 - 0.8
\,{\rm GeV}\,{\rm cm^{-3}}$) (See \citet{bss83, cal81, turner86,
  bergstrom98}).  The range quoted above is from \citet{bergstrom98}
where a variety of halo profiles were considered.

We find (Table 3) $\rho_{\rm local} = 0.0080\pm0.0014\,M_\odot\,{\rm
  pc}^{-3}$ and $\sigma_{\rm local} = 240\pm 23\,{\rm km\,s^{-1}}$.
The PDFs for these two quantities are plotted in Figure \ref{fig:dm}.
Our values for the mean and lower bound for $\rho_{\rm local}$ are
consistent with the values found in \citet{bergstrom98} though our
analysis suggests that their upper bound is too high by a factor of
two.  Our mean value for $\sigma_{\rm local}$ is lower than the
standard value by about $30\,{\rm km\,s^{-1}}$ though the standard
value still falls within the range of acceptable models.  Our models
tend to favour values for $M_{100}$ at the low end of the range found
in equation \ref{eq:mass100}.  Inspection of a scatter plot of models
in the $M_{100}-\sigma_{\rm local}$ plane reveals a clear correlation
between the two quantities --- models with values of $M_{100}$ closer
to the central in equation \ref{eq:mass100} have values of
$\sigma_{\rm local}$ closer to the standard value.

\subsection{\it Connection with Cosmology}

\citet{kly02} construct models for the Milky Way based on standard
galaxy formation theory.  In particular, they use model halos based on
predictions from cosmological simulations with the further assumption
that the halos undergo adiabatic contraction in response to the
formation of the disk and bulge.

Though our MCMC analysis does not include cosmological constraints we
can test whether our models are consistent with the standard
cosmological paradigm {\it a posteriori}.  To this end, we calculate
the virial radius, $R_{\rm vir}$, virial mass, $M_{\rm vir}$, and
concentration, $c_{\rm vir}$, for all of the models in our MCMC series.
By definition, the mean density inside $R_{\rm vir}$ is a factor
$\Delta_{\rm vir}$ greater than the background density, $\rho_m$.  That is,
\begin{equation}
M_{\rm vir}\equiv M\left (R_{\rm vir}\right )\equiv
\frac{4\pi}{3}\Delta_{\rm vir}\rho_m R_{\rm vir}^3
\end{equation}
where $M_{\rm vir}$ is the mass interior to $R_{\rm vir}$.
$\Delta_{\rm vir}$ depends on the cosmological model; In what follows,
we assume $\Omega_m\equiv \rho_m/\rho_{\rm crit} = 0.3$ where
$\rho_{\rm crit}\equiv 3H^2/8\pi$ is the critical density and $H =
70\,{\rm km \,s^{-1}\,Mpc}$ \citep{tegmark}.  With these values,
$\Delta_{\rm vir}\simeq 340$ \citep{bryan98} (see also
\citet{bullock01} and \citet{wechsler02}).  The concentration, $c_{\rm
  vir}$, is defined as the ratio of $R_{\rm vir}$ to the halo scale
length $R_s$, the latter identified as the radius at which the
logarithmic slope of the halo density profile equals -2. For a halo
profile given by equation \ref{eq:haloprofile}, $R_s = \left
  (2-\gamma\right )a_h$.

In Figure 12, we show the PDF for our MCMC series projected onto the
$M_{\rm vir}-c_{\rm vir}$ and $R_{\rm vir}-c_{\rm vir}$ planes.  Also
shown are the low-concentration, high-concentration, and favored
models from \citet{kly02}.  In the upper panels, $R_{\rm vir}$,
$M_{\rm vir}$, and $c_{\rm vir}$ are calculated for the actual halos
used in our models.  The implication would seem to be that our models
are more concentrated than the ones assumed in \citet{kly02}.

Recall, however, that \citet{kly02} incorporate adiabatic contraction
into their models.  In order to compare our halos with theirs, we
should assume they too have undergone adiabatic contraction.  We
should therefore adiabatically ``de-contract'' our halos and then
calculate $c_{\rm vir}$.  We have done this using the usual
assumptions \citep{blum86, flores93}; The system is treated as being
spherically symmetric.  Initially, the baryons and dark matter are
well-mixed.  The baryons cool and form a disk and bulge while the halo
particles respond adiabatically to the changing gravitational
potential.  Moreover, halo particles are assumed to follow circular
orbits which do not cross as their orbits shrink.  Under these
assumptions, the quantity $rM(r)$ remains constant and one can calculate
the initial radii of the dark matter particles given the final structure
of the disk, bulge, and halo.

  The results of this analysis are shown in the lower panels of Figure
  12.  Our models are now in excellent agreement with those found in
  \citet{kly02}.

  \citet{alam02} proposed $\Delta_{V/2}$, the mean density within the
  radius $R_{V/2}$ in units of $\rho_{\rm crit}$, as a measure of the
  halo core densities.  $R_{V/2}$ is defined as the
  radius at which the rotation curve reaches half its maximum value
  $V_{\rm max}$.  The quantity $\left (\Delta_{V/2}/8\pi^2\right
  )^{1/2}$ is equal to the number of rotation periods per Hubble time
  at the radius $R_{V/2}$.  \citet{alam02} found values of
  $\Delta_{V/2}$ in the range $5\times 10^4-5\times 10^6$ for dark
  matter-dominated galaxies, considerable scatter and no correlation
  between $\Delta_{V/2}$ and $V_{\rm max}$.  

  The PDF for our MCMC series in the $\Delta_{V/2}-V_{\rm max}$ plane
  is shown in Figure 13.  The range of values for $\Delta_{V/2}$ is
  certainly consistent with those found in the \citet{alam02} survey
  though some models have higher values --- possibly reflecting that
  influence of baryons on the Milky Way's dark halo.  The considerable
  range indicates that the halo concentration is poorly constrained by
  the data used in our analysis.

\section{BAR FORMATION}

Near-IR photometry, gas and stellar kinematic measurements, and
observations of gravitational microlensing events all strongly suggest
that the Milky Way is a barred galaxy (see reviews by \citet{kui96}
and \citet{ger96}).  A bar represents a strong departure from
axisymmetry and adds considerably to the challenge of modeling the
Galaxy.  A promising avenue is to use N-body simulations to follow
the development of bars and spiral structure in an initially
axisymmetric, equilibrium model.  \citet{ost73} and \citet{sel85}
provide early examples of this approach.  They were interested in
stabilizing their Galactic models to avoid bar formation, the former
proposing an unseen dark halo and the latter stressing the importance
of the bulge.  More recently, \citet{fux97} attempted to construct
self-consistent models for the Milky Way's bar by evolving unstable
axisymmetric models and comparing the results with observations from
the DIRBE experiment \citep{dwe95}.

Bars, at least in idealized, initially-axisymmetric disk galaxies,
form through swing amplification \citep{toomre81}.  (Whether bars in
real galaxies form through this process or through some more
complicated process during the formation of the galaxy itself is
another matter.)  In swing amplification, leading waves propagate
outward and are amplified into trailing waves (and contained by the
outer Lindblad resonance).  Trailing waves then wind up.  Within
linear theory, if there is an inner Lindblad resonance (ILR), it
absorbs the tightly wound trailing waves and thereby acts as a barrier
preventing further growth.  In the absence of an ILR, the trailing
waves propagate through the center of the system and transform into
leading waves.  This feedback loop can lead to growth of a bar-like
perturbation.

The Toomre $Q$ \citep{toomre64} and Goldreich-Tremaine $X$
\citep{gt78, gt79} parameters are the two most widely used diagnostic
quantities for studying a galaxy's resistance to the bar instability.
$Q$ measures the kinetic ``temperature'' of the disk; stellar disks
with $Q<1$ are unstable to local gravitational instabilities.  $X$
indicates a disk's susceptibility to global instabilities.  ($X$
depends on the azimuthal mode number, $m$, of the perturbation.  Here,
we take $m=2$ since we are interested in bars.)  For disks with $X\la
3$, the gain of the swing amplifier is large and bars are more likely
to form.  In general, the greater the contribution of the disk to the
gravitational force felt by disk particles, the smaller the value of
$X$.

Higher values of $Q$ and $X$ make a galaxy more resistant to the bar
instability.  The existence of an ILR barrier from a dense bulge or
cuspy halo can also prevent the instability \citep{sel01}.  However, a
bar can still form even if the galaxy initially has an ILR barrier.
For example, interactions between halo substructure and the bulge or
cusp may temporarily lower the ILR barrier and trigger bar formation
in an otherwise stable galaxy \citep{jr06}.  Non-axisymmetric
structure in the disk might also jostle the cusp and remove the ILR
barrier, if only temporarily.  In short, the notion of an ILR barrier
assumes linear perturbation theory; non-linear disturbances may be
able to overcome or disrupt the barrier and initiate the bar instability.

Our models provide a natural starting point for investigations of bar
formation.  The models generated by our MCMC run all yield acceptable
fits to the observational data yet have very different stability
properties.  Figure 14 shows a contour plot of the model distribution
in the $Q-X$ plane.  (Of course, both $Q$ and $X$ are functions of
radius.  Here we use their minimum values.)  Also shown is the
distribution of models in the $X'-X$ plane where
\begin{equation}
\label{eq:xprime}
X'\equiv \frac{v^2_{\rm total}(R)}{v^2_{\rm disk}(R)}_{R = 2.2R_d}
\end{equation}
is a measure of the disk's contribution to the gravitational force
necessary to keep a particle in a circular orbit at a given radius.
Here $v_{\rm total}$ is the circular rotation speed at cylindrical
radius $R$ and $v_{\rm disk}$ is the contribution to $v_{\rm total}$
from the disk \citep{deb98, deb00}.  The radius $R=2.2R_d$ is where
$v_{\rm disk}$ reaches a maximum, assuming an exponential disk.  The
tight correlation between $X$ and $X'$ indicates these two quantities
are interchangeable for most purposes.

We consider twenty-five models from our MCMC run which cover most of
the area within the 95\% likelihood contour in the $Q-X$ plane.  The
models in this study are depicted as dots in Figure \ref{fig:qxx}
while their circular-speed curves are shown in Figure \ref{fig:vgrid}.
In all cases, the circular speed reaches a peak value of approximately
$220\,{\rm km\,s^{-1}}$ at a radius between $5-8\,{\rm kpc}$ and then
declines slowly.

Figure \ref{fig:lindblad} shows the behavior of the functions $\Omega$
and $\Omega\pm \kappa/2$ where $\Omega$ is the angular velocity and
$\kappa$ is the epicyclic radial frequency.  These functions should be
compared with the pattern speed, $\Omega_p$, of an emerging bar or
spiral density wave.  In particular, $\Omega = \Omega_p$ indicates the
position of co-rotation while $\Omega_p = \Omega - \kappa/2$ indicates
the position of the ILR.  As we will see shortly, all of the bars that
form in our simulations have initial pattern speeds in the range of
$\sim 40-50\,{\rm km\,s^{-1}\,kpc^{-1}}$.  Thus, about two thirds of the
models in this study have initial ILRs.  Note that in these cases, the
rise in the $\Omega - \kappa/2$-curve is very rapid and the radius of
the ILR is only a few hundred parsecs.

Each of the twenty-five models depicted in Figure \ref{fig:qxx} are
evolved for $5\,{\rm Gyr}$ using a parallelized treecode
\citep{dub96}.  Simulations have 800K disk, 200K bulge, and 1M halo
particles.  The particle Plummer softening length is $\epsilon=25$ pc
and the simulations are run for $10^4$ equal timesteps of length
$\Delta t=0.5$ Myr.  We generate surface density maps for the face on
view of every tenth timestep and determine the strength and pattern
speed of the bars that form.  An animation depicting the evolution of
the grid of simulations is available at
http://www.cita.utoronto.ca/$\sim$dubinski/DynamicalBlueprints/

In Figure \ref{fig:ia2} we plot the amplitude of the bar-strength
parameter
\begin{equation}\label{eq:m2mode}
A_2 \equiv \frac{1}{N}\left |\sum_j e^{2i\phi_j}\right |~,
\end{equation}
as a function of time.  Here $\phi_j$ is the usual azimuthal angle of
the $j^{\rm th}$ disk particle and $N$ is the total number of disk
particles.  All of the models form bars, with the possible exception
of the $Q=2,~X=4.5$ model (upper right-hand corner of the figures).
Bars form almost immediately in models with low values of $Q$ and $X$.
Indeed, the models in the lower left-hand region of the figures are
violently unstable and radically transformed by both local and global
instabilities.  On the other hand, bars can take several billion years
to form in models with higher values of $Q$ and especially higher
values of $X$.

Interestingly enough, the growth-rate of bars does not vary smoothly
across the grid of models.  In particular, models in the central
column of the grid ($Q\simeq 1.5$) with higher values of $X$ quickly
develop bars.  In general, these models have less massive and less
dense bulges and no ILRs (see Figures 15 and 16).  Hence, they are
immediately vulnerable to Toomre's swing amplification mechanism.

Virtually all of the models with ILRs eventually form bars though the
onset of the instability is typically delayed.  Two effects, one
numerical and one physical, can lead to bar formation in the presense
of an initial ILR.  First, two-body relaxation may be
important at radii near the ILR even with $10^6$ halo particles.
(Recall that in some models, the radius of the ILR can be as small as
$100\,{\rm pc}$.)  As the central halo relaxes, the peak of the
$\Omega-\kappa/2$-curve can drop below the initial pattern speed.  If
this occurs, the ILR barrier disappears and the bar forms.

A second, more physical, explanation is that swing amplified spiral
waves disturb the central region of the galaxy causing the ILR barrier
to disappear, at least temporarily.  Other non-linear disturbances,
such as interactions between the disk and halo substructure, may also
cause the ILR barrier to disappear.  Simply put, the theory of the ILR
barrier may not apply to situations where non-linear perturbations are
present.  We note that the importance of nonlinear effects for bar
formation is discussed in \citet{sell89}.

The evolution of the bar length, $R_b$, and pattern speed, $\Omega_p$,
are shown in Figures \ref{fig:bar_evolution} and \ref{fig:omega}.  To
estimate the bar length, we first construct the axis ratio profile of
the isodensity contours for the disk.  The axis ratio profile is
defined to be the axis ratio, $b/a$, as a function of the semi-majar
axis, $a$.  In general, $b/a$ goes through a minimum with a typical
value of $(b/a)_{\rm min}\simeq 0.4$ before rising abruptly in the
transition from the end of the bar to the outer disk.  As a heuristic
measure of $R_b$ we use the major axis length of the isodensity
contour beyond the $b/a$-minimum and where $b/a\simeq 0.6$.  The
pattern speed is given by the time-derivative of the phase angle
$\phi_0=\tan^{-1} [{\rm Im}(A_2)/Re(A_2)]/2$.  We insure that $\phi_0$
is sampled with enough time-resolution to avoid aliasing and to obtain
a reasonable estimate of the pattern speed from $\Delta \phi/\Delta
t$.  In Figure \ref{fig:omega}, we show the pattern speed once the bar
is easily detected above the noise.

Though the bars in our survey appear at different times, their initial
pattern speeds are always in the range of $40-50\,{\rm
  km\,s^{-1}\,kpc^{-1}}$.  Angular momentum transfer to the halo
causes $\Omega_p$ to decay to about $20\,{\rm km\,s^{-1}\,kpc^{-1}}$
over a few Gyr with the decay rate being slightly larger in models
with higher mass halos.  As expected, the bar lengths are generally
less than the co-rotation radius though transient long bars with $R_b
\approx 10$ kpc form in models with small $Q$ and $X$ before
collapsing on themselves.  For most models, bars grow monotonically in
length as the pattern speed declines.

Bar formation alters the structure of the model and it is therefore
natural to ask whether the evolved models still satisfy the original
observational constraints.  Though we will leave the details of such
an analysis for future work, we include, in Figure
\ref{fig:denevol}, the evolution of the surface density profiles for
our twenty-five simulated systems.  We see that the surface density
profiles of models with low values of $Q$ and $X$ are dramatically
deformed; the redistribution of mass is so violent that the models
almost certainly do not satisfy our observational constraints.  On the
other hand, models where the bar forms relatively late in the
simulation show little evolution of the surface density profile.
Their structure, at least in an azimuthally-averaged sense, remains
largely unchanged.

The Galaxy is known to have a bar and estimates of its length and
pattern speed can be compared with our results.  \citet{bin97} find
$R_b=3.5$ kpc and $\Omega_p \approx 60-70\,{\rm
  km\,s^{-1}\,kpc}^{-1}$ in their analysis of DIRBE photometry while
\citet{deh99} find a similar bar length but lower pattern speed
($\Omega_p= 53\pm 3 \,{\rm km\,s^{-1}\,kpc^{-1}}$) using the velocity
distribution of solar neighborhood stars.  \citet{wei99} model the gas
kinematics of the inner galaxy and find $\Omega_p \approx 42 \,{\rm
  km\,s^{-1}\,kpc^{-1}}$.  The fact that nearly all of the models in
our study are bar unstable and have initial pattern speeds near the
range of the inferred values is promising.  We note that certain
models can be excluded such as those with very small values of $Q$ and
$X$.  If we assume that the Galaxy's bar has formed very recently than
many of the models have the correct combination of bar length and
pattern speed to match the observations.  Furthermore, for larger $Q$
and $X$, the change in the disk's radial profile in response to the
bar is small.  These models may well provide a good barred model of
the Galaxy.  It should be noted that in all of these models, the
pattern speed declines to $20-30 \,{\rm km\,s^{-1}\,kpc^{-1}}$ within a
few Gyr after the bar forms. {\em If we take these models seriously as
  reasonable facsimiles of the Galaxy, then we must conclude that
  Galactic bar formed within the last 1-2 Gyr.}

\section{CONCLUSIONS}

We have introduced a dynamical model for late-type galaxies that
incorporates our current understanding of disk-bulge-halo systems.  In
particular, the bulge has a Sersic surface density profile and the
halo has a central density cusp. 

We have carried out an MCMC analysis of dynamical models for the Milky
Way using a variety of kinematic and photometric constraints.  The
results are presented in the form of PDFs for both input parameters
and derived quantities.  The MCMC analysis provides a picture of the
distribution of models in parameter space that is more complete than
can be obtained by other approaches.  Avoided is the awkward procedure
of fixing a subset of parameters while allowing the remaining
parameters to vary in some minimization scheme.  Instead, a sequence
of models is generated which contains all of the desired information.
Marginalization over a subset of parameters is accomplished by simply
projecting the model distribution onto the appropriate parameter
subspace.

Our analysis suggests that the Milky Way has a pseudo-bulge with a
Sersic index of $1.3\pm 0.3$.  Our results for the masses of the disk,
bulge, and halo are consistent with those of \citet{deh98} but call
into question choices for these quantities in some popular models from
the literature.  For example, the disk and bulge masses in
\citet{johnston99} are entirely inconsistent with our results.  The
inferred bulge mass for the standard model used by the MACHO
collaboration \citep{alcock00} is inconsistent with our findings by a
factor of 2.5.  On the other hand, the standard values used by
terrestrial dark matter detection experiments for the local dark
matter density and velocity dispersion are consistent with our
results.

A weak point of our analysis is the inability to tightly constrain the
halo mass at large radii.  Planetary nebulae, globular clusters, and
satellite galaxies may be used as tracers of the Galactic potential.
Dynamical models for the tracer populations are required to properly
model kinematic data.  In principle, it is straightforward to
construct such models but there are subtle issues.  Previous
studies showed that velocity anisotropy in a tracer population can
affect interpretation of kinematic data.  Velocity anisotropy requires
a DF that depends on at least one integral of motion in addition to
the energy.  The standard practice is to use the total angular
momentum, $L$, but since our models include a disk, $L$ is not
conserved.  (Previous analyses side-stepped this issue by using a
spherically symmetric Galactic potential.)

Our MCMC analysis provides an ideal starting point for studies of disk
stability and bar formation in that we have some $10^5$ models, each
of which can serve as initial conditions for a numerical experiment.
We have performed a suite of simulations which focuses on the
susceptibility of the disk to bar formation as a function of the
stability parameters $Q$ and $X$.  Many of the models provide a good
match to the inferred properties of the Galactic bar with the proviso
that the bar has formed recently.

\acknowledgements{It is a pleasure to thank S. Courteau, S. Dodelson,
  R. de Jong, A. Graham, P. Gregory, S. Myers, D. Puglielli, J. Sellwood,
  D. Spergel, and A. Toomre for useful conversations.  This work was
  supported by the Natural Sciences and Engineering Research Council
  of Canada.  Simulations were conducted on facilities funded by the
  Canadian Foundation for Innovation at the Canadian Institute for
  Theoretical Astrophysics.}


\clearpage

\clearpage

\clearpage

\begin{figure}
\epsscale{1.0}
\plotone{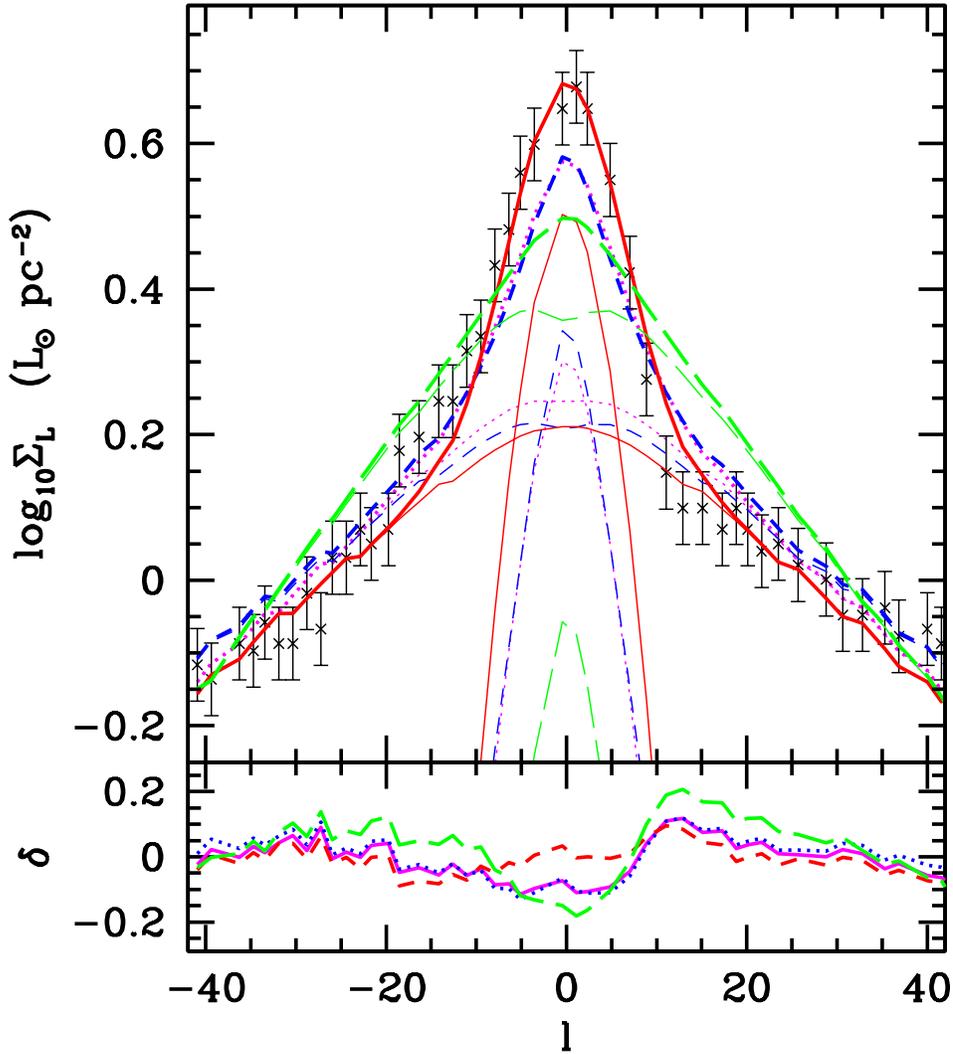}
\label{fig:sbp}
\caption{Surface brightness profile as a function of $l$ for four
  models and for data from \citet{spe96}.  Three models are chosen
  from our main MCMC run: $\left (n,\,\gamma\right ) \simeq \left
    (0.6,1\right )$ --- red curve; $\left (n,\,\gamma\right ) \simeq
  \left (1,1\right )$ --- magenta curve; $\left (n,\,\gamma\right )
  \simeq \left (2,1\right )$ --- blue curve.  Also shown (green curve)
  is a model chosen from a run where $n$ is fixed to the ``de
  Vaucouleurs'' value, $4$ and $\gamma=0.86$.  In the top panel, the
  thin curves show the separate contributions of the disk and bulge.
  Lower panel shows the residuals between the models and the data.}
\end{figure}

\clearpage

\begin{figure}
\epsscale{1.0}
\plotone{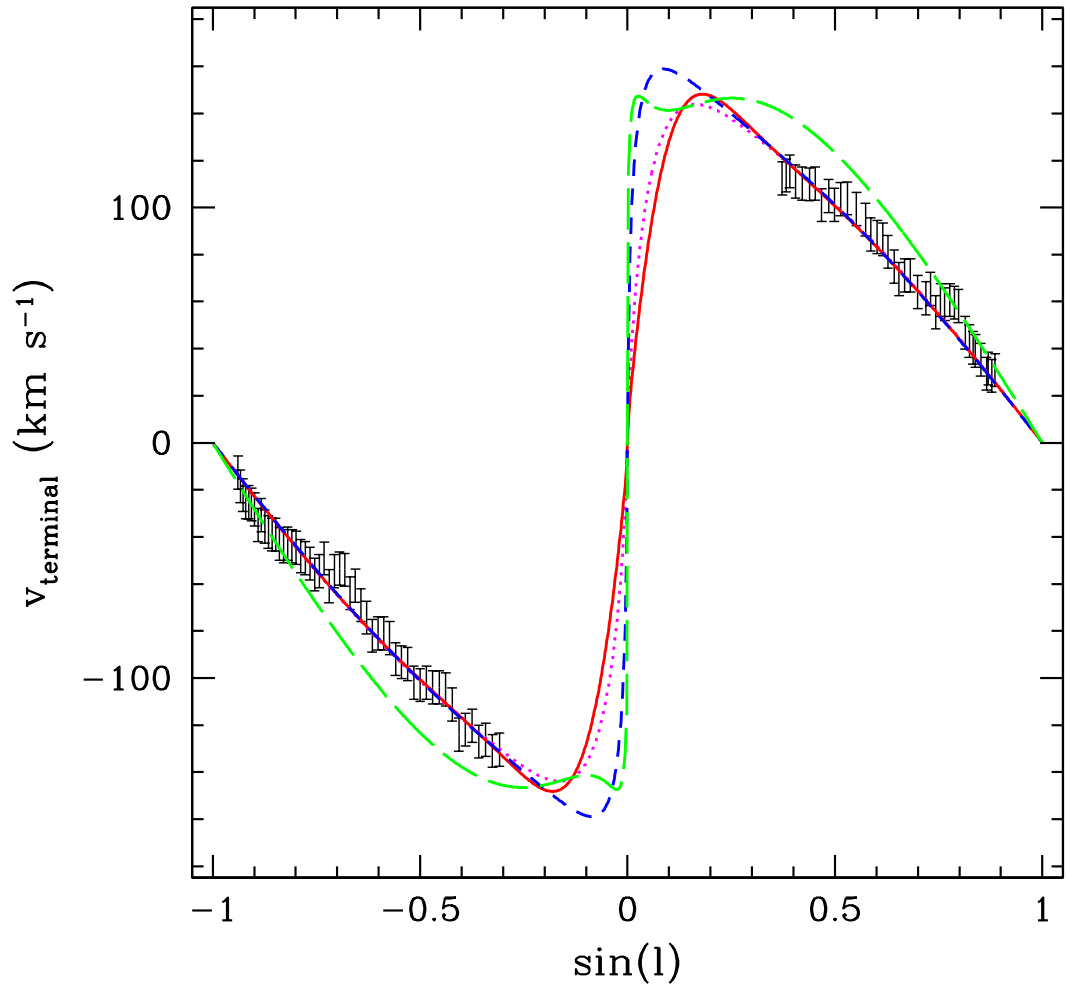}\label{fig:vterminal}
\caption{Terminal velocity as a function of $\sin{l}$ for four models
  and for data from \citet{mal95} Line types and colours are the same
  as in Figure 4.}
\end{figure}

\clearpage

\begin{figure}
\epsscale{1.0}
\plotone{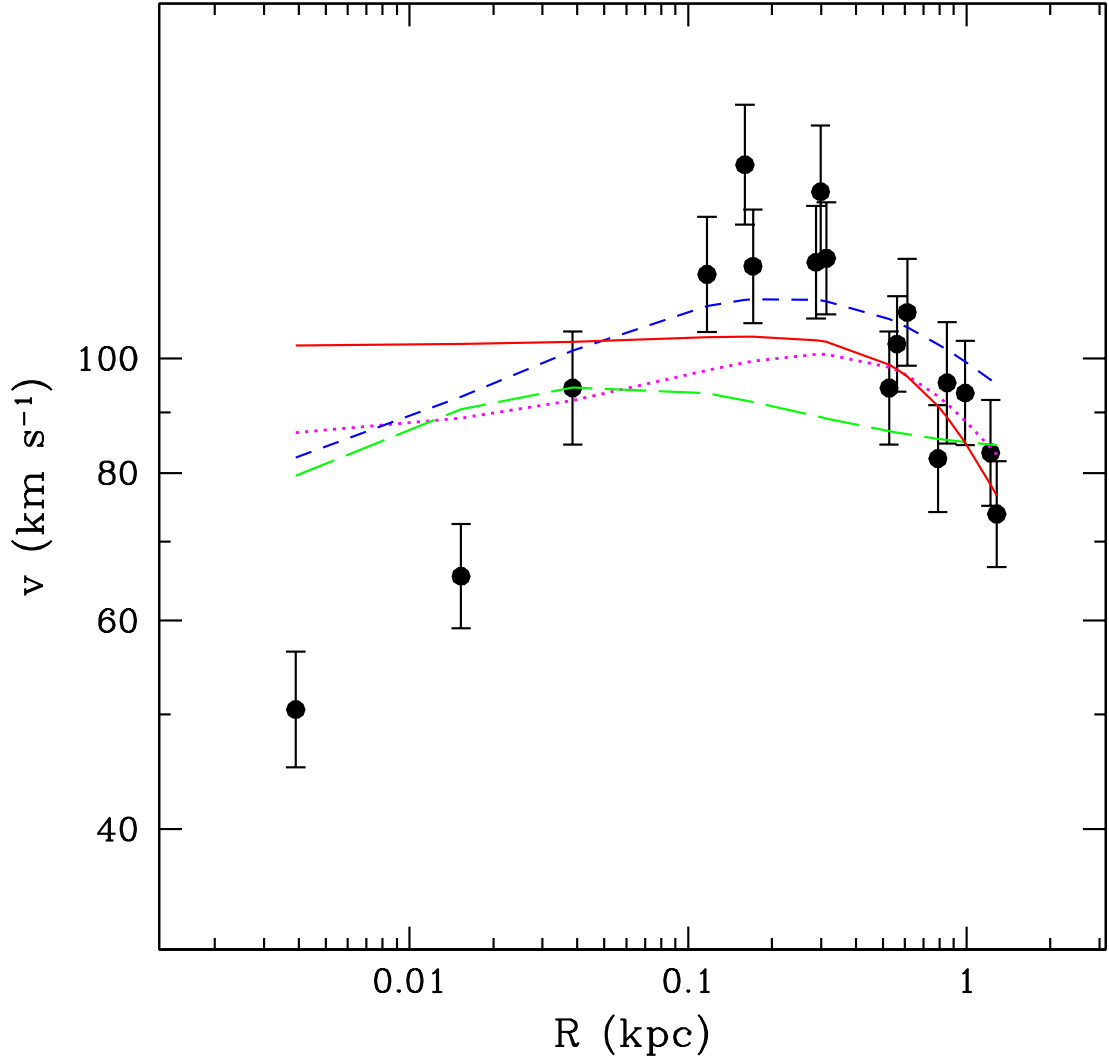}
\label{fig:vlos}
\caption{Line-of-sight velocity dispersion toward the bulge as a
function of projected radius from the Galactic center for 
models and for data from \citet{tre02}.  Line types and colours 
are the same as in Figure 4.}
\end{figure}

\begin{figure}
\epsscale{1.0}
\plotone{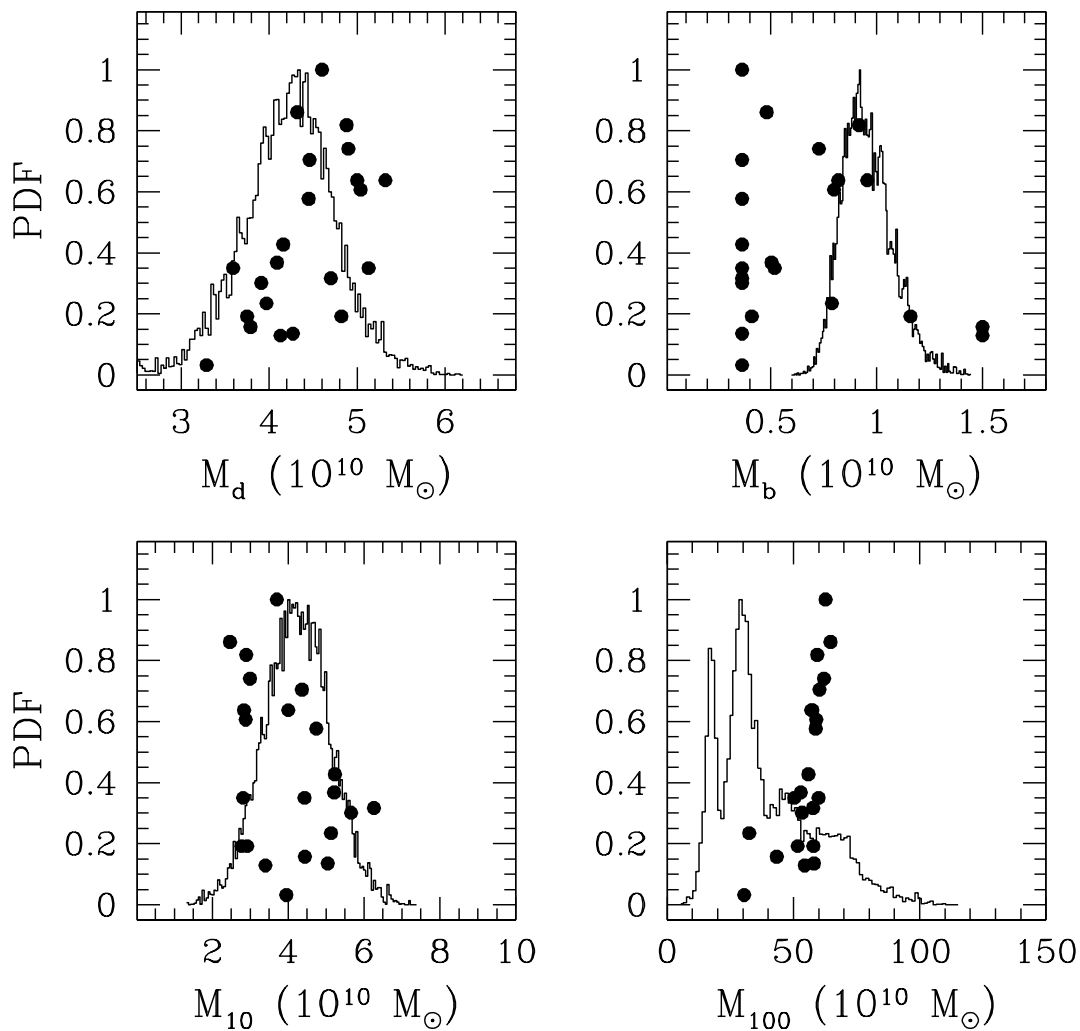}
\caption{PDFs for the disk mass, bulge mass and halo mass within
$10$ and $100$ kpc.  Solid lines show results of the MCMC analysis.
Dots show likelihood functions for the twenty-two models presented by
\citet{deh98}.}
\label{fig:masspdfs}
\end{figure}

\begin{figure}
\epsscale{1.0}
\plotone{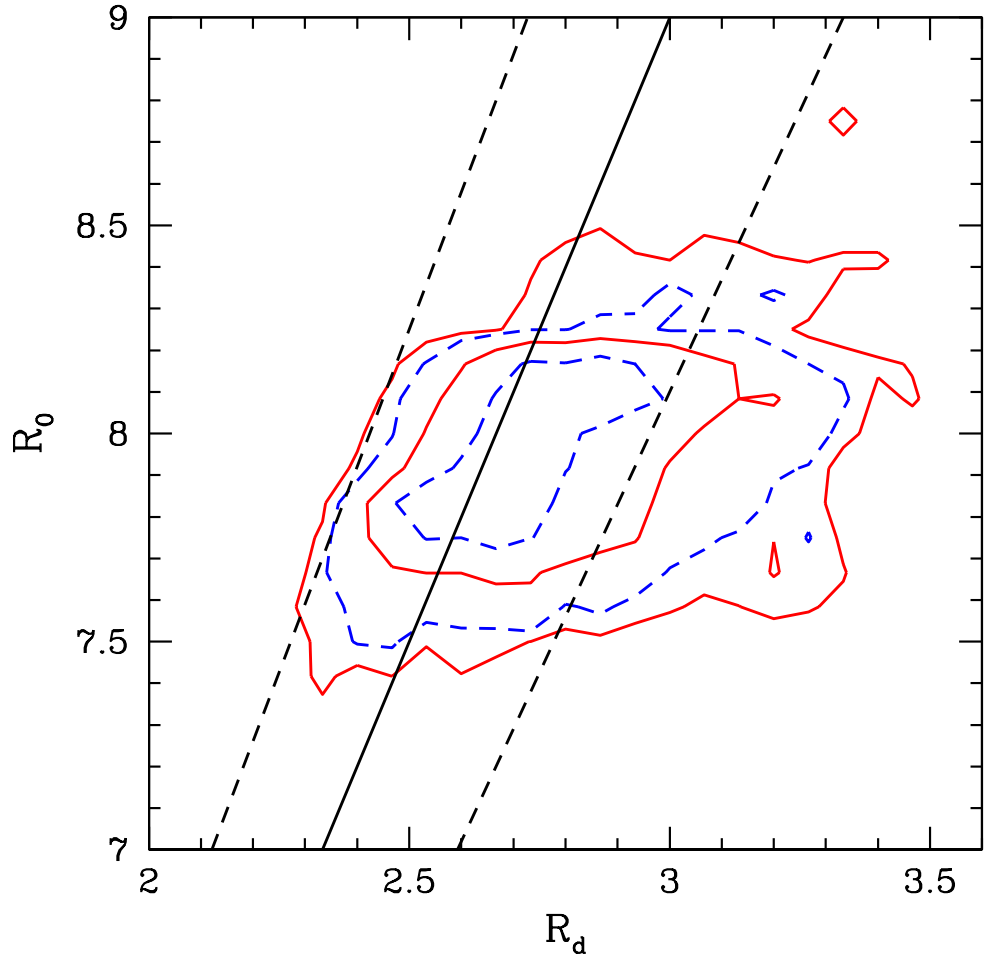}
\caption{Contour plot of probability distribution function of models
  in the $R_0-R_d$ plane.  Solid contours enclose $68\%$ and $95\%$ of
  the models.  Dashed contours enclose $38\%$ and $87\%$ of the
  models.  The solid straight (black) line corresponds to $R_0/R_d=3$.
  The dashed straight lines correspond to $R_0/R_d=2.7~{\rm and}~3.3$.}
\label{fig:ro_rd}
\end{figure}

\begin{figure}
\epsscale{1.0}
\plotone{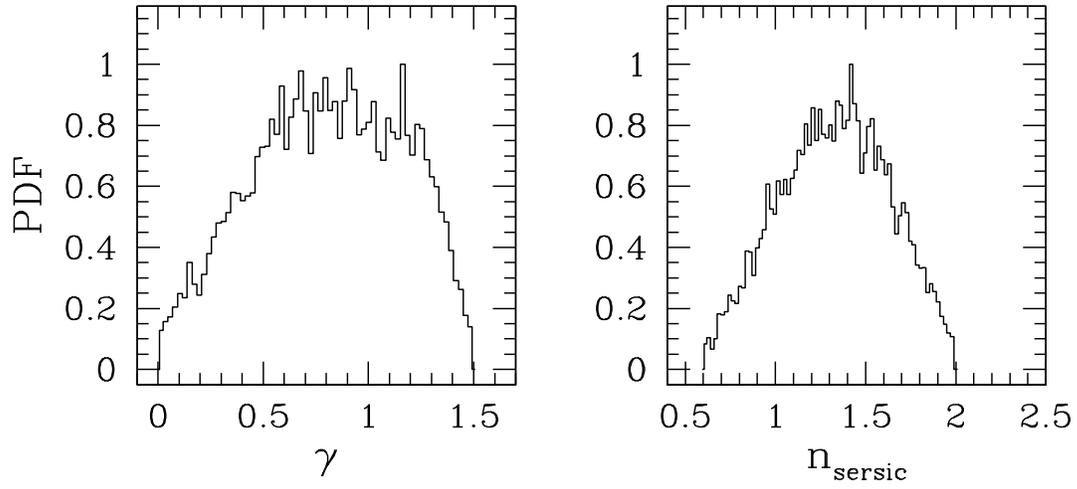}
\label{fig:sersicgamma}
\caption{PDFs for the Sersic index, $n$, and cusp-strength parameter,
$\gamma$.}
\end{figure}

\begin{figure}
\epsscale{0.9}
\plotone{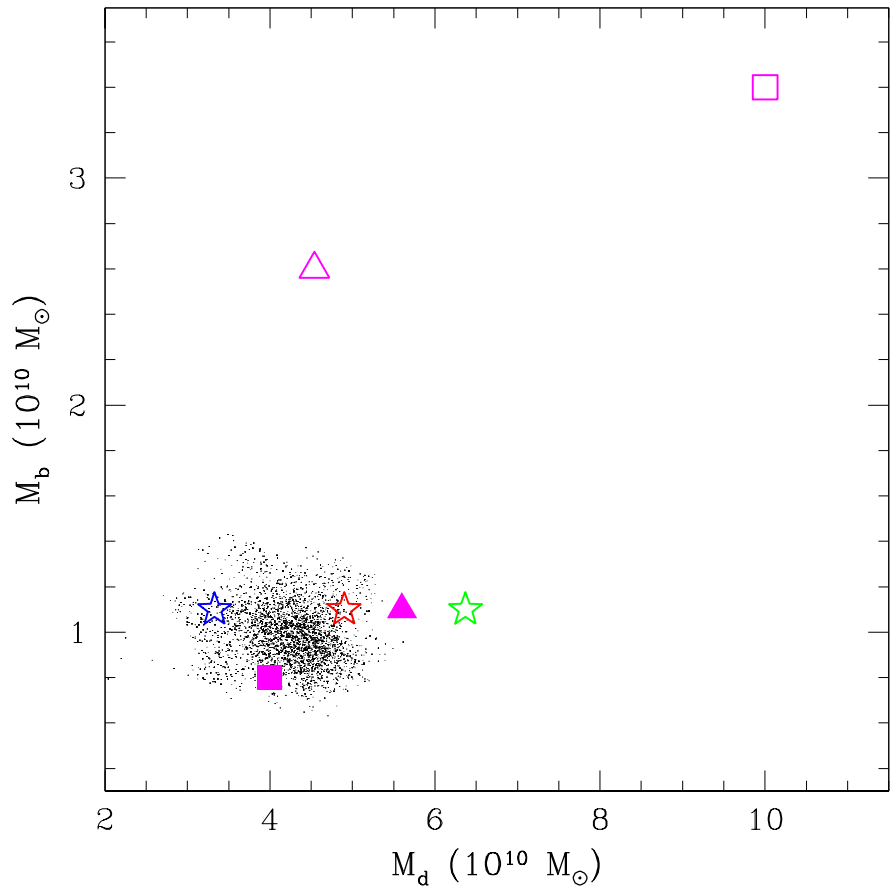}
\caption{Comparison of our results for the disk and bulge mass with
  those from published Milky Way models.  Cloud of points in the lower
  right corner are from the MCMC analysis.  Stars from left to right
  (blue, red, green) represent, respectively, the low, high, and
  maximal models from \citet{ken92}.  The filled triangle represents
  the model from \citet{bss83} while the filled square represents the
  model from \citet{kly02}.  The open square represents the model
  advocated by \citet{johnston99} in their studies of the tidal
  disruption of Sagittarius.  The open triangle is the standard model
  adopted by the MACHO collaboration.}
\label{fig:modelcomparison}
\end{figure}

\begin{figure}
\epsscale{1.0}
\plotone{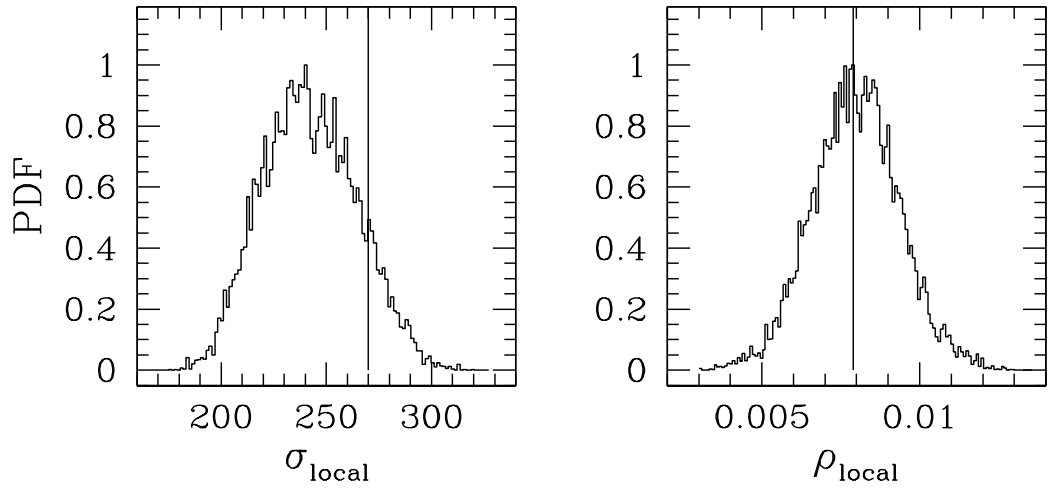}
\label{fig:dm}
\caption{PDFs for the local dark matter velocity dispersion,
  $\sigma_{\rm local}$ and dark matter density, $\rho_{\rm local}$.
  Vertical lines indicate the standard values assumed by most
  terrestrial dark matter detection experiments (see text).}
\end{figure}

\begin{figure}
\epsscale{1.0}
\plotone{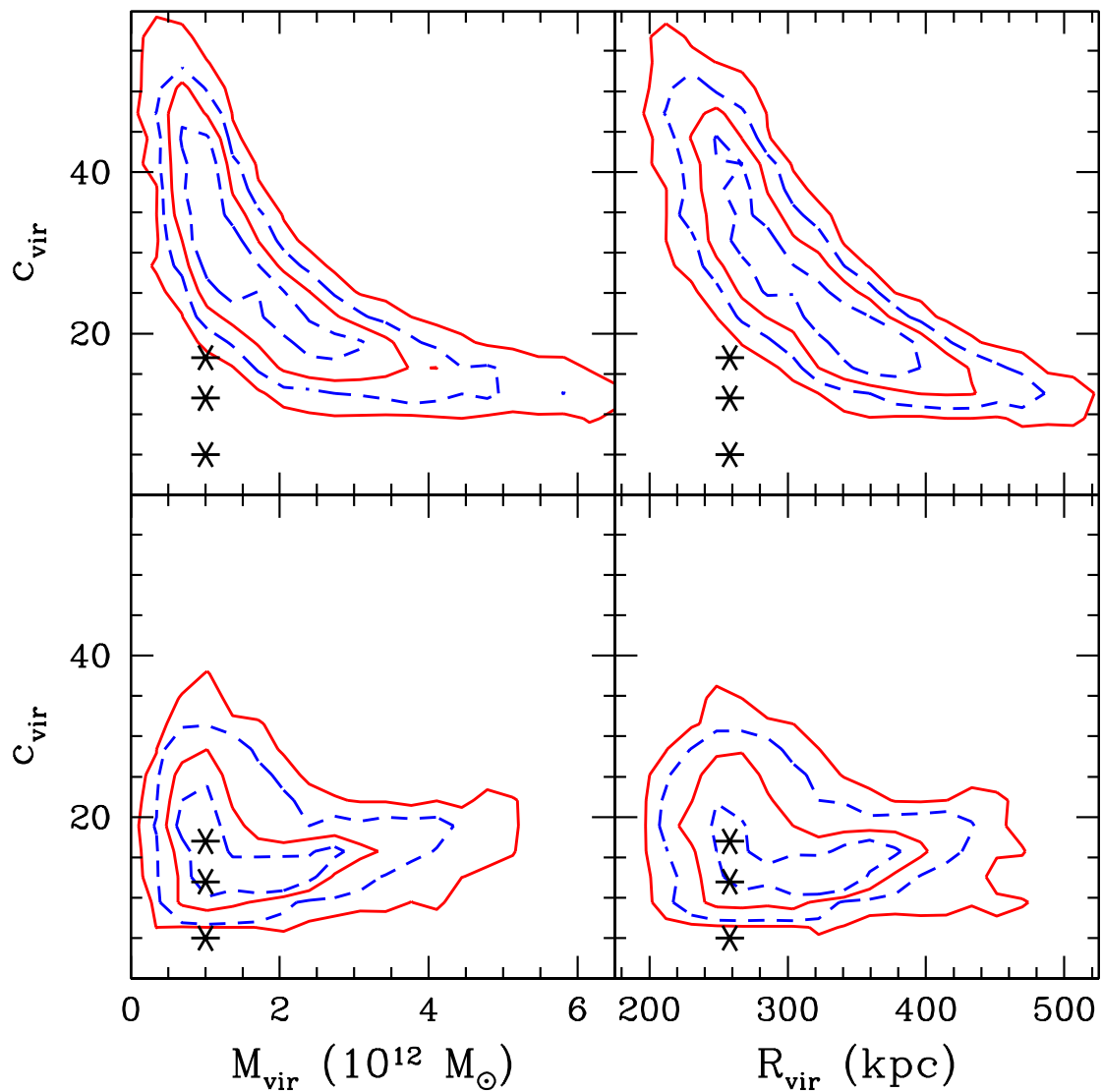}
\caption{Contour plots of the probability distribution function of
models in the $M_{\rm vir}-c_{\rm vir}$ and $R_{\rm vir}-c_{\rm vir}$ planes.
Solid contours enclose $68\%$ and $95\%$ of the models.  Stars indicate
the favored, low-concentration ($c_{\rm vir} = 5$) and high-concentration
($c_{\rm vir}=17$) models from \citet{kly02}.}
\label{fig:cvir}
\end{figure}

\begin{figure}
\epsscale{1.0}
\plotone{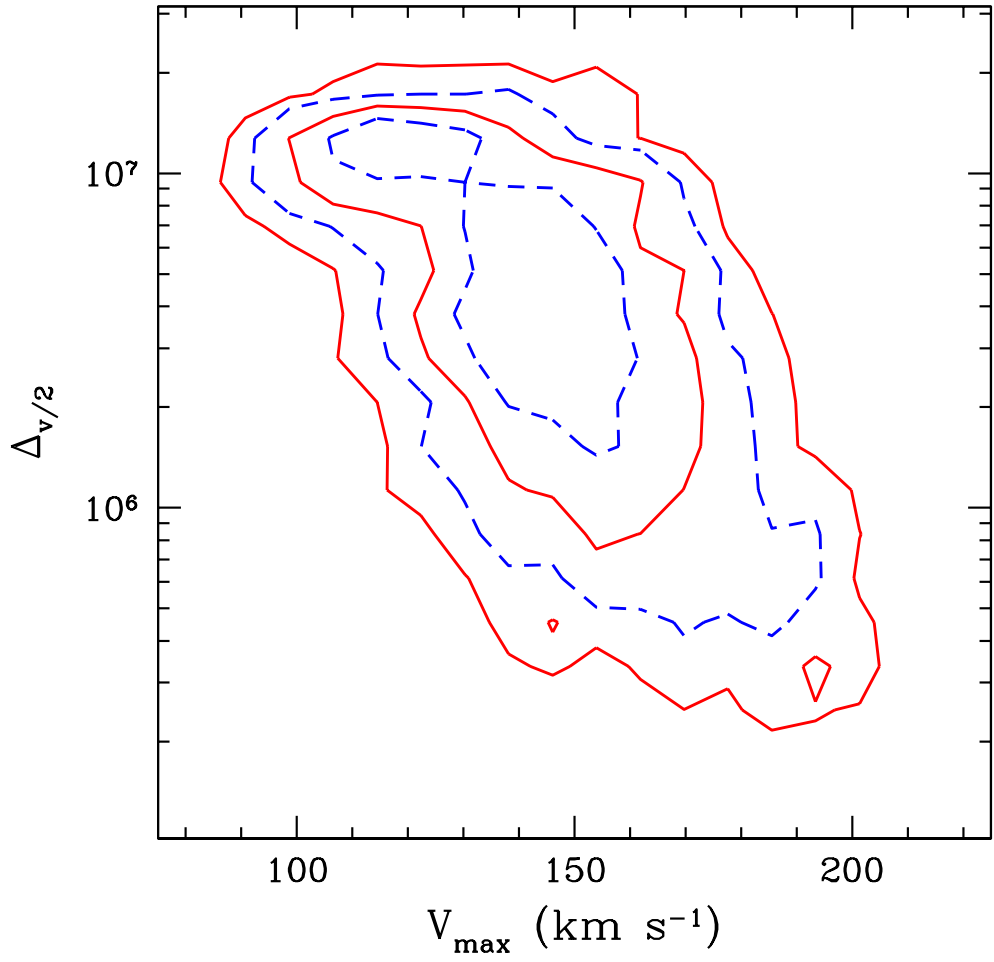}
\caption{Contour plot of the probability distribution function 
of $\Delta_{V/2}-V_{\rm max}$.}
\label{fig:deltav2}
\end{figure}

\begin{figure}
\epsscale{1.0}
\plotone{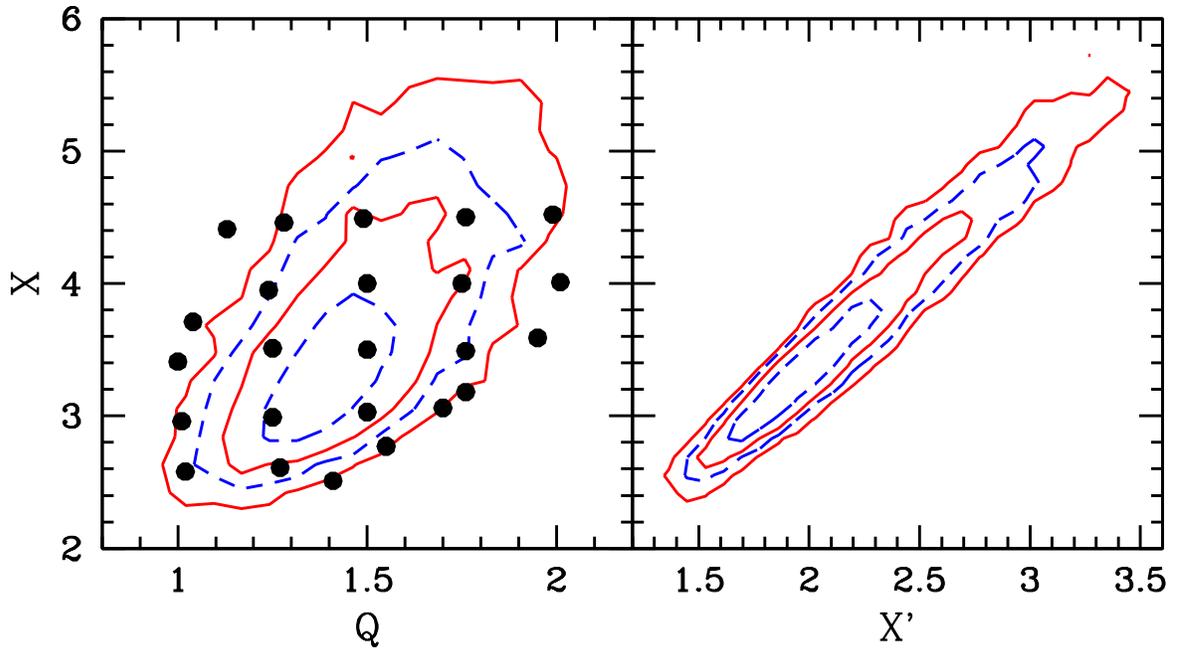}
\caption{Contour plots of probability distribution function of models
  in the $Q-X$ and $X'-X$ planes.  Solid contours enclose $68\%$ and
  $95\%$ of the models.  Dots correspond to models used in bar
  formation study in Section 7.}
\label{fig:qxx}
\end{figure}

\begin{figure}
\epsscale{1.0}
\plotone{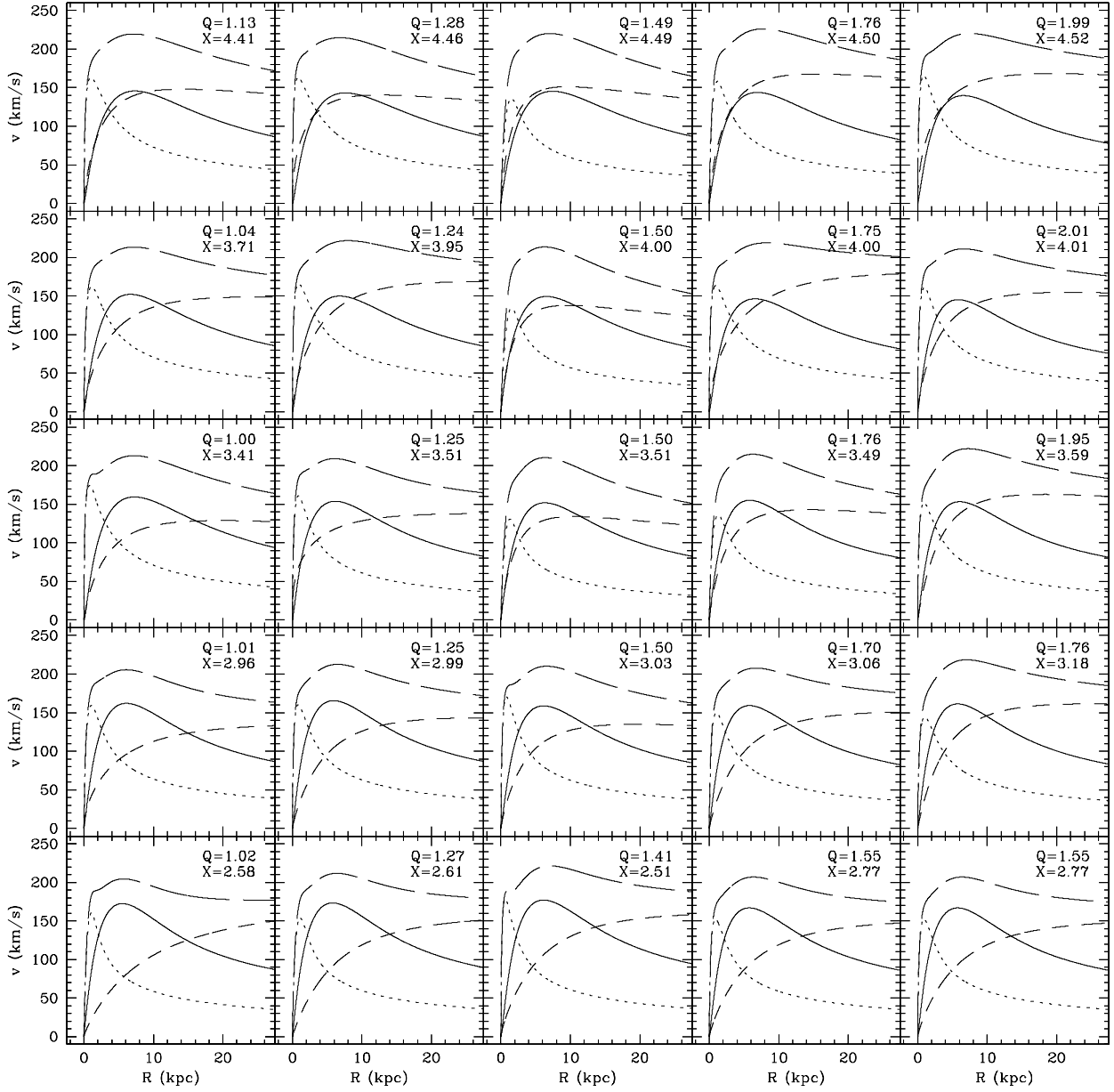}
\caption{Rotation curves for the twenty-five models used in our bar
  formation simulations.  Models are arranged so that $Q$ increases to
  the right and $X$ increases from bottom to top.  Shown are the total
  rotation curve (long-dashed line) and contributions to the rotation
  curve from the disk (solid curve), bulge (dotted curve), and halo
  (dashed curve).}
\label{fig:vgrid}
\end{figure}

\begin{figure}
\epsscale{1.0}
\plotone{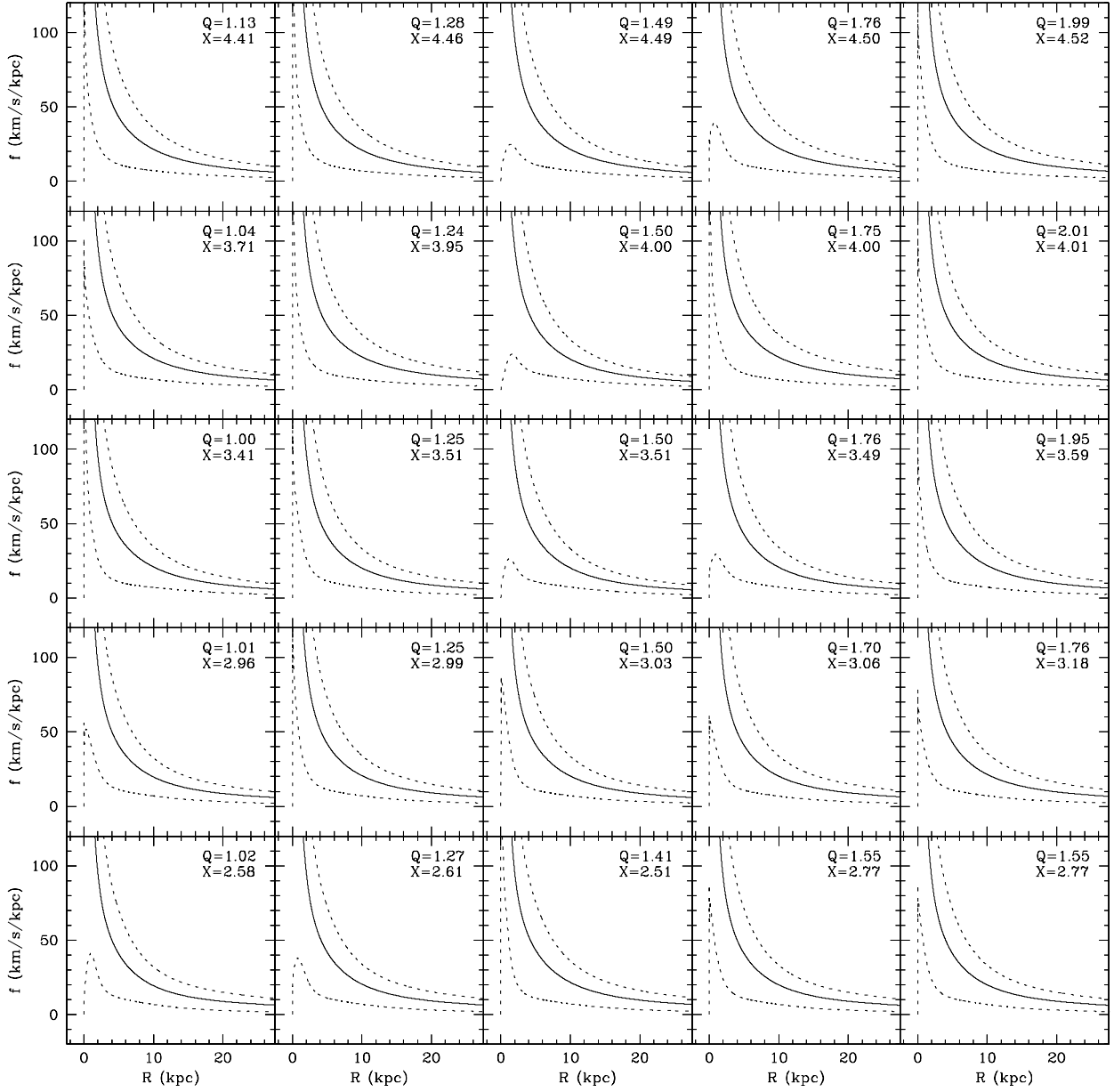}
\caption{$\Omega-n\kappa/m$ as a function of radius for 
$n=0$ (solid curve), $n=1,\,m=\pm 2$ (dotted curves).}
\label{fig:lindblad}
\end{figure}

\begin{figure}
\epsscale{1.0}
\plotone{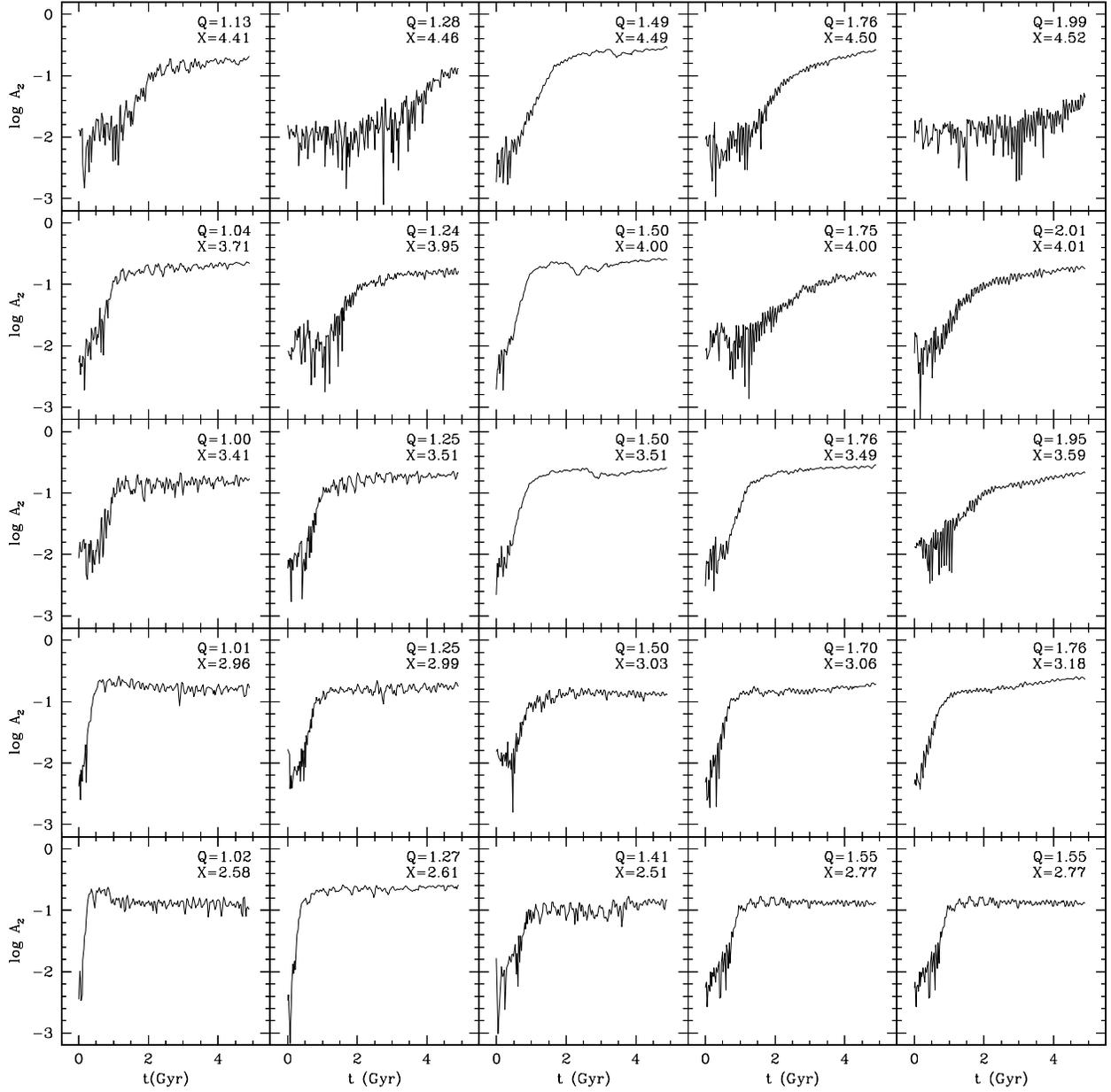}
\caption{Growth of the bar strength parameter, $A_2$, as a function of time.}
\label{fig:ia2}
\end{figure}

\clearpage

\begin{figure}
\epsscale{1.0}
\plotone{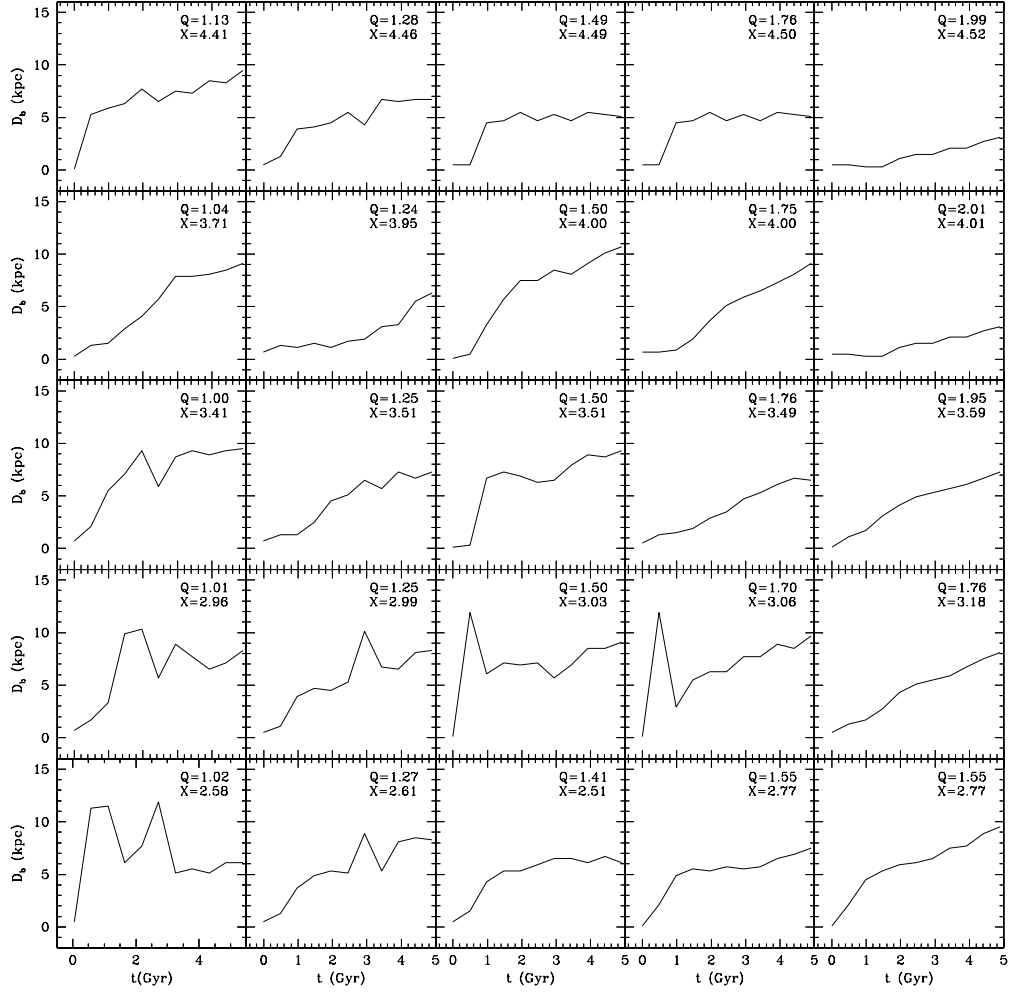}
\caption{Bar length, $R_b$ as a function of time.}
\label{fig:bar_evolution}
\end{figure}

\begin{figure}
\epsscale{1.0}
\plotone{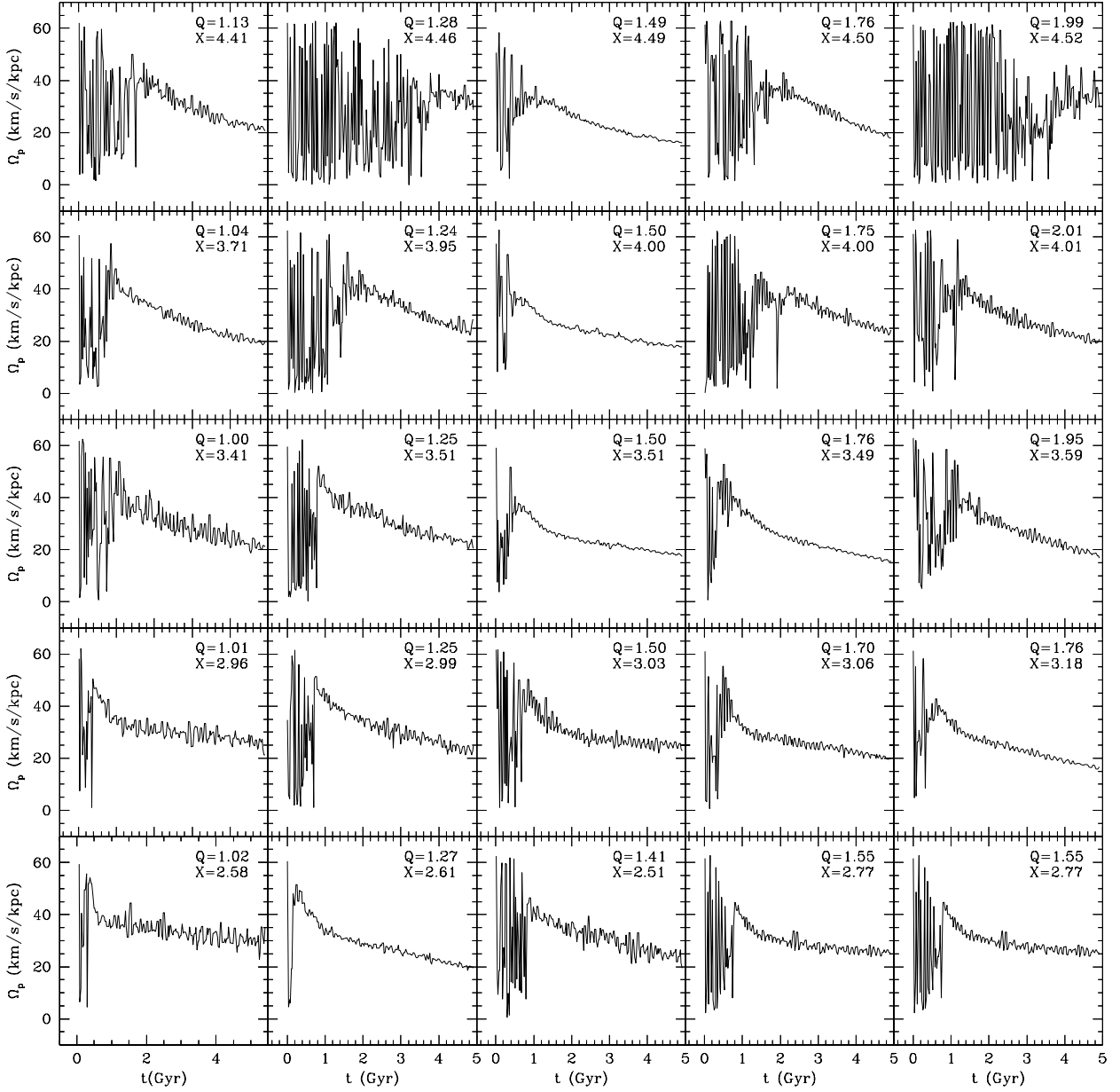}
\caption{The bar pattern speed $\Omega_p$ as a function of time.  Bars are
born with pattern speeds $\Omega_p \sim 50\,{\rm km}\,{\rm s}^{-1}\,{\rm
kpc}^{-1}$ which immediately begin to decay as they transfer angular
momentum to halos.}
\label{fig:omega}
\end{figure}

\begin{figure}
\epsscale{1.0}
\plotone{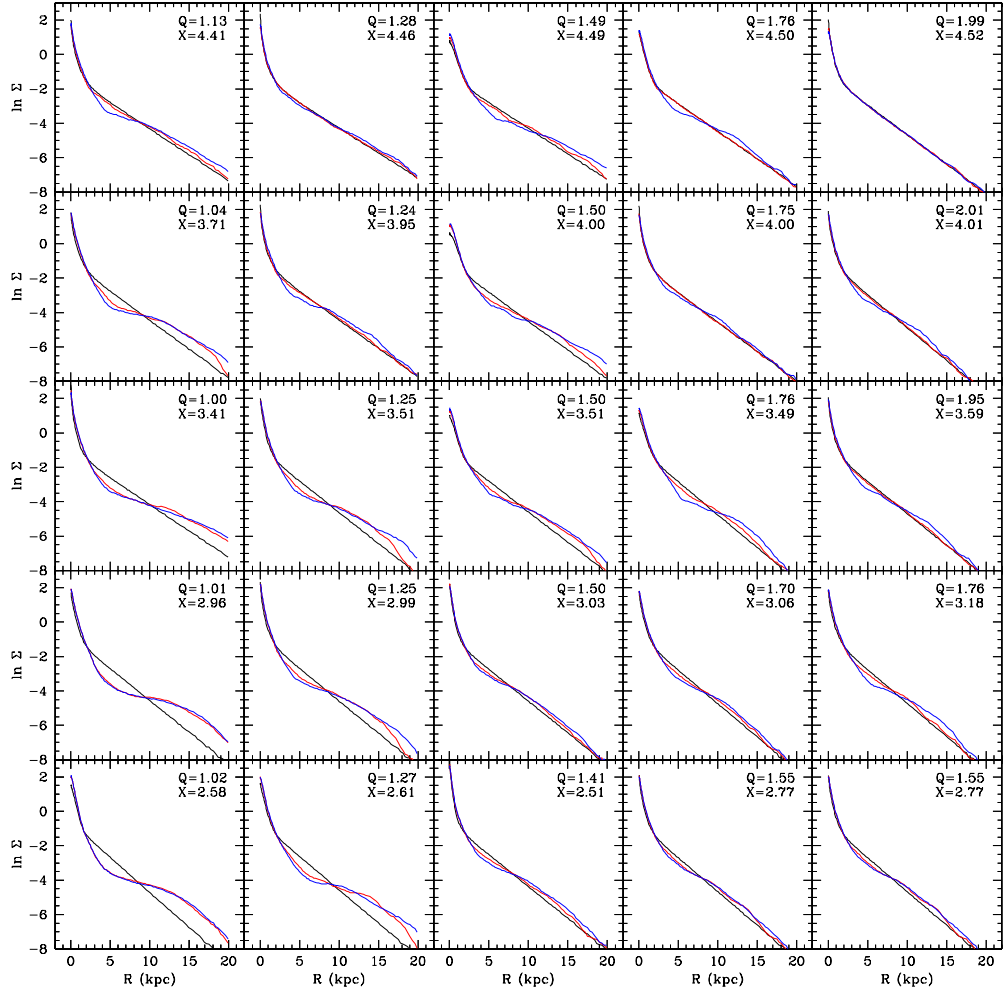}
\caption{Surface density profile evolution.  Three times are shown: the
initial time $t=0$ (black),  $t=2.5$ Gyr (red), and $t=5.0$ Gyr
(blue).}
\label{fig:denevol}
\end{figure}

\begin{deluxetable}{cccc}
\tablewidth{0pt}
\tabletypesize{\footnotesize}
\tablecaption{model parameters and prior probabilities}
\tablehead{ \colhead{parameter} & {prior} & {lower bound} & {upper bound} \\}
\startdata
$\sigma_h$ &  Jeffreys & $2\, {\rm km\,s^{-1}}$ & $6\, {\rm km\,s^{-1}}$\\
$a_h$	&  Jeffreys  & $2\, {\rm kpc}$ & $35\, {\rm kpc}$\\
$\gamma$ & uniform & 0 & 1.5 \\
$M_d$ &  Jeffreys & $2\times 10^{10} M_\odot$ & $ 7\times 10^{10} M_\odot$\\
$R_d$ &  Jeffreys & 2 $ {\rm kpc}$ & 3.5 ${\rm kpc}$\\
$h_d$ &  Jeffreys & 0.2 ${\rm kpc}$ & 1 $ {\rm kpc}$\\
$\sigma_{R0}$ &  Jeffreys & 80 ${\rm km\,s^{-1}}$ & 300 ${\rm km\,s^{-1}}$\\
$n$ &  uniform & 0.6 & 2.0\\
$\sigma_b$ &  Jeffreys & 150 ${\rm km\,s^{-1}}$ & 400 ${\rm km\,s^{-1}}$\\
$R_e$ &   Jeffreys  & 0.4 ${\rm kpc}$ & 1 ${\rm kpc}$\\
$\left (M/L\right )_d$ & uniform & 0.6 & 1.2 \\
$\left (M/L\right )_b$ & uniform & 0.6 & 1.2 \\
$R_0$ & Jeffreys & 7 {\rm kpc} & 9 {\rm kpc} \\
\enddata
\label{tab:table1}
\end{deluxetable}

\begin{deluxetable}{ccc}
\tablewidth{0pt}
\tabletypesize{\footnotesize}
\tablecaption{results for input parameters}
\tablehead{ \colhead{parameter} & {average} \\}
\startdata
$\sigma_h$ &  $  330_{-35}^{+35}$ \\
$a_h$	&   $  13.6_{-9.0}^{+12.2}$ \\
$\gamma$ & $0.81_{0.39}^{+0.39}$\\
$M_d$ &    $  4.1_{-0.5}^{+0.53}$ \\
$R_d$ &   $  2.8_{-0.22}^{+0.23}$\\
$h_d$ &  $  0.36_{-0.04}^{+0.04}$ \\
$\sigma_{R0}$ &  $  119_{-13}^{+13}$ \\
$n$ & $1.32_{0.33}^{0.32}$ \\
$\sigma_b$ &   $272_{-25}^{+25}$ \\
$R_e$ &  $  0.64_{-0.09}^{+0.09}$ \\
$\left (M/L\right )_d$ & $0.96_{0.09}^{0.1}$\\
$\left (M/L\right )_b$ & $0.60_{0.06}^{0.07}$\\
$R_0$ & $7.94_{0.2}^{0.2}$ \\
\enddata
\label{tab:table2}
\tablecomments{Units are: ${\rm km\,s}^{-1}$ for velocities,
$10^{10}\,M_\odot$ for masses, and ${\rm kpc}$ for distances.}
\end{deluxetable}

\begin{deluxetable}{ccc}
\tablewidth{0pt}
\tabletypesize{\footnotesize}
\tablecaption{results for calculated quantities}
\tablehead{ \colhead{parameter} & {average} \\}
\startdata
$M_d$ & $ 4.22_{-0.50}^{+0.52}$ \\
$M_b$ & $ 0.96_{-0.12}^{+0.12}$ \\
$M_{10}$& $ 4.23_{-0.86}^{+0.88}$ \\
$M_{25}$& $ 12.6_{-2.8}^{+2.9}$ \\
$M_{50}$& $ 24_{-8.27}^{+9.4}$\\
$M_{100}$ & $ 40.0_{-19}^{+22}$ \\
$\rho_0$& $ 0.0080_{-0.0014}^{+0.0014}$ \\
$\sigma_0$ & $ 241_{-23}^{+23}$ \\
\enddata
\label{tab:table3}
\tablecomments{Units for $\rho_0$ are $M_\odot\,{\rm pc}^{-3}$.  Other
quantities use the same units as in Table 2.}
\end{deluxetable}

\end{document}